\documentclass{acmart}

\usepackage{comment}
\usepackage[utf8]{inputenc}
\usepackage{comment}

\usepackage{amsmath,amsthm,amsfonts,mathtools,mathrsfs,bbm,bm}
\usepackage[first=0,counter=rand]{lcg}
\usepackage{pdfpages}
\usepackage{tabularx}
\usepackage{algorithm}
\usepackage{caption}
\usepackage[noend]{algpseudocode}
\usepackage{marginnote}
\usepackage{float}
\usepackage[hypcap=false]{caption}
\usepackage{subcaption}
\usepackage{tikz}
\usetikzlibrary{shapes,arrows,matrix,positioning,fit,quotes,decorations.pathreplacing,calc}

\usepackage[toc,page]{appendix}
\usepackage{listings}
\usepackage{todonotes}

\usepackage{ulem} 

\DeclareMathAlphabet{\mathbx}{U}{BOONDOX-ds}{m}{n}
\SetMathAlphabet{\mathbx}{bold}{U}{BOONDOX-ds}{b}{n}
\DeclareMathAlphabet{\mathbbx} {U}{BOONDOX-ds}{b}{n}

\algnewcommand{\LeftComment}[1]{\Statex \(\triangleright\) #1}

\newcommand{\tikzmark}[1]{\tikz[overlay,remember picture] \node (#1) {};}

\newcommand*{\SpaceReservedForComments}{1.7cm}%
\newcommand*{\HorizontalOffset}{-0.5em}%
\newcommand*{\VerticalOffset}{0.7ex}%
\newcommand*{\AddNote}[4][]{%
    \begin{tikzpicture}[overlay, remember picture]
        \draw [decoration={brace,amplitude=0.5em},decorate,ultra thick,red, #1]
            ($(#3)+(\HorizontalOffset,-\VerticalOffset)$) --  ($(#2)+(\HorizontalOffset,\VerticalOffset)$)
            node [align=left, text width=\SpaceReservedForComments-1.0em, pos=0.5, anchor=east] {#4};
    \end{tikzpicture}
}

\newcommand*{\AddNoteOneLine}[4][]{%
    \begin{tikzpicture}[overlay, remember picture]
        \draw [decoration={brace,amplitude=0.2em},decorate,ultra thick,red, #1]
            ($(#3)+(\HorizontalOffset,-\VerticalOffset)$) --  ($(#2)+(\HorizontalOffset,\VerticalOffset)$)
            node [align=left, text width=\SpaceReservedForComments-1.0em, pos=0.5, anchor=east] {#4};
    \end{tikzpicture}
}

\makeatletter
    \algrenewcommand\alglinenumber[1]{\tikzmark{\arabic{ALG@line}}\tiny#1:}
\makeatother

\makeatletter
\newcommand{\multiline}[1]{%
  \begin{tabularx}{\dimexpr\linewidth-\ALG@thistlm}[t]{@{}X@{}}
    #1
  \end{tabularx}
}
\makeatother

\DeclarePairedDelimiter{\ceil}{\lceil}{\rceil}

\algnewcommand\algorithmicto{\textbf{to}}
\algrenewtext{For}[3]{%
	\algorithmicfor\ #1 \textbf{from} #2 \algorithmicto\ #3 \algorithmicdo
}
\algnewcommand\COMMENT[2][.25\linewidth]{%
	\leavevmode\hfill\makebox[#1][l]{$\triangleright$~#2}
}
\algnewcommand\LCOMMENT[2][.4\linewidth]{%
	\leavevmode\hfill\makebox[#1][l]{$\triangleright$~#2}
}
\algnewcommand\RETURN{\State \textbf{return} }

\algnewcommand{\algorithmicgoto}{\textbf{go to}}%
\algnewcommand{\GoTo}[1]{\algorithmicgoto~\ref{#1}}%

\algrenewcommand\algorithmicprocedure{$\triangleright$}

\newcommand{\dd}{\mathop{}\,\mathrm{d}}

\newsavebox{\smlcommat}
\savebox{\smlcommat}{$\left(\begin{smallmatrix}4 & 4\end{smallmatrix}\right)$}
\newsavebox{\smlconmat}
\savebox{\smlconmat}{$\left(\begin{smallmatrix}1/4 & 1/8\\0.0 & 1/2\end{smallmatrix}\right)$}

\newcommand{\sched}{\text{Sched}}

\tikzset{%
	highlight/.style={rectangle,rounded corners,fill=red!15,draw,fill opacity=0.5,thick,inner sep=0pt}
}

\graphicspath{{./images/}}

\tikzset{pics/.cd,
	graph/.style args={#1}{
	code = {
		\ifnum #1 = 1
		\begin{scope}[all/.style={fill=orange!20},all,every node/.append style={all,draw,circle}]
			\node [draw, circle] (N0) at (0,0) {$1$};
			\node [draw, circle] (N3) at (6,-1) {$4$};
			\node [draw, circle] (N4) at (0,-3) {$5$};
			\node [draw, circle] (N6) at (6,-3) {$7$};
		\end{scope}
		\else
		\begin{scope}[all/.style={},all,every node/.append style={all,draw,circle}]
			\node [draw, circle] (N0) at (0,0) {$1$};
			\node [draw, circle] (N3) at (6,-1) {$4$};
			\node [draw, circle] (N4) at (0,-3) {$5$};
			\node [draw, circle] (N6) at (6,-3) {$7$};
		\end{scope}
		\fi
		\begin{scope}[all/.style={draw, circle, fill=orange!50},all,every node/.append style={all}]
		\path (2,-1) node [fill=red!70] (N1) {$2$};
		\path (4,-1) node (N2) {$3$};
		\path (3,-3) node (N5) {$6$};
		\end{scope}

		\ifnum #1 = 1
		\node[draw=black, double, dashed, fit=(N0) (N1) (N2) (N3) (N4) (N5) (N6), ellipse] (tmp) {};
		\else
		\node[draw=black, double, dashed, fit=(N1) (N2) (N5), ellipse] (tmp) {};
		\fi

		\draw [-latex] (N0) edge (N1);
		\draw [-latex] (N1) edge (N2);
		\draw [-latex] (N1) edge (N5);
		\draw [-latex] (N2) edge (N5);
		\draw [-latex] (N3) edge[bend right] (N0);
		\draw [-latex] (N4) edge (N1);
		\draw [-latex] (N3) edge (N5);
		\draw [-latex] (N2) edge (N3);
		\draw [-latex] (N3) edge (N6);
		\draw [-latex] (N6) edge (N5);
	}
	}
}

\newcommand{\uca}{Universit\'e C\^ote d'Azur}

\acmJournal{TOMACS}

\title[Efficient Simulation of Sparse Graphs of Point Processes]{
Efficient Simulation of Sparse Graphs of Point Processes
}

\author{Cyrille Mascart}
\orcid{0000-0002-8600-9545}
\affiliation{%
    \position{Phd. student}
    \institution{\uca}
    \department{Laboratoire I3S}
    \streetaddress{2400 route des Lucioles}
    \city{Biot}
    \postcode{06000}
    \country{France}
}
\email{mascart@i3s.unice.fr}
\author{Alexandre Muzy}
\affiliation{\uca, CNRS, I3S,  France}
\email{alexandre.muzy@univ-cotedazur.fr}
\author{Patricia Reynaud-Bouret}
\affiliation{\uca, CNRS, LJAD,  France}
\email{patricia.reynaud-bouret@univ-cotedazur.fr}

\date{\today}

\keywords{Point processes, discrete event simulation, Hawkes point processes, computational complexity, local independent graphs}

\begin{document}
    \begin{abstract}
        We derive new discrete event simulation algorithms for marked time point processes. The main idea is to couple a special structure, namely the associated local independence graph, as defined by Didelez~\cite{didelez}, with the activity tracking algorithm~\cite{cise-muzy} for achieving high performance asynchronous simulations. With respect to classical algorithm, this allows reducing drastically the computational complexity, especially when the graph is sparse.
    \end{abstract}

\ccsdesc[500]{Mathematics of computing~Stochastic processes}
\ccsdesc[500]{Computing methodologies~Discrete-event simulation}
\ccsdesc[300]{General and reference~Performance}
\ccsdesc[100]{Applied computing~Bioinformatics}
    \maketitle
    \clearpage
    \tableofcontents
    \clearpage
    
	
	\section{Introduction}
        Point processes in time are stochastic objects that model efficiently event occurrences. The variety of applications is huge: from medical data applications (time of death or illnesses) to social sciences (dates of crimes, weddings, etc), from seismology (earthquake occurrences) to micro-finance (actions of selling or buying a certain assets), from genomics (gene positions on the DNA strand) to reliability analysis (breakdowns of complex systems) (see e.g. \cite{andersen, didelez, Hoff, VeO82, RBS, cha}). 
        
        Most of the time, point processes are multivariate, in the sense that either several processes are considered at the same time, or in the sense that one process regroups together all the events of the different processes and marks them by their type. A typical example consists in considering either two processes, one counting the wedding events of a given person and one counting the birth dates of children of the same person. One can see this as a marked process which regroups all the possible dates of birth or weddings independently and on each event one marks it by its type, here wedding or birth.
        
        In the sequel, we denote the individual process $N_j$, the set of all events corresponding to type $j$, for $j=1,...,M$ and the joint process $N=N_1\cup .. \cup N_m$. In this multivariate or marked case, the individual processes are usually globally dependent, the apparition of one event or point on a given type influencing the apparition of other points for the other types and the simulation of the whole system cannot be easily parallelized.
        
        This is especially true in neuroscience \cite{IEEE}. Let us detail a bit more this set up which is a benchmark example here. Neurons are excitable electric cells that are linked together inside a huge network ($10^{11}$ for humans~\cite{nbNeurons}, $10^8$ for rats~\cite{Herculano-Houzel2518}, $10^6$ for coackroaches), each cell receives inputs from approximately $10^3$ to $10^4$ presynaptic (upstream) neurons~\cite{PAKKENBERG200395}. Depending on its excitation, the neuron might then produce an action potential also called spike, information which is propagated to postsynaptic (downstream) neurons.
        
        From a stochastic point of view, one might then see the spike trains emitted by a given neuron as an individual point process which in fact is embedded in a multivariate point process with $M$, the total number of neurons as the total number of types. The size of the network requires then very well adapted simulation schemes that may use the relative sparseness of the network with respect to the global size of the network.
        
        To do so, we use the mathematical notion of local independence graph for marked point processes due to Didelez \cite{didelez}, which is detailed in Section \ref{dist} and which informally corresponds to the real neuronal network. In this sense, \emph{in the sequel we call marks, processes and type the nodes of the graph}.
         The only strong assumption that is used is the \emph{time asynchrony hypothesis}, (i.e. points or events of different mark or types, meanings points or events appearing in different nodes, cannot occur at the exact same time) together with the fact that all processes have a \emph{conditional intensity} \cite{bremaud1981point}.
        
        Simulation of point processes has a long history that dates back to Doob in the 40's \cite{Doob} for Markov processes. In the 70's, Gillespie \cite{gillespie} popularized the  method for a particular application: chemical reactions. At the same time,  Lewis and Shedler \cite{lewis} proposed a thinning algorithm for  simulating univariate inhomogeneous Poisson processes (this can also be viewed as a rejection method). Few years later, Ogata \cite{ogata} produced a hybrid algorithm able to  simulate multivariate point processes in the general case even if they are not markovian, including both a choice of the next point by thinning and a choice of the node to activate thanks to Gillespie principle. This method is still up to now the benchmark for simulating such processes, and is for instance used in recent packages such as \texttt{ppstat} in \texttt{R} (2012). It has been rediscovered many times in various cases, most of the time as a Generalized Gillespie  method (see for instance \cite{barrio}).
        
        When the number of types or nodes is huge, this method can quickly become inefficient in terms of computational times. Many people have found shortcuts, especially in Markovian settings. For instance, Peters and de With \cite{peters} proposed  a new simulation scheme exploiting  a network of interaction for particular physical applications. In \cite{doucet}, the authors reformulated this algorithm in a more mathematical way for a particular case of Piecewise Deterministic Markov Processes. People have even exploited very particular structures such as  Hawkes processes, with exponential interactions (special case which leads to Markovian intensities) \cite{dassios}, to be able to simulate huge networks, as in the \texttt{Python} package \texttt{tick} (2017).
        
        In the mean time, the technique of discrete event simulation first appeared in the mid-1950s~\cite{tocher} and was used to simulate the components (machines) of a system changing state only at discrete ``events''. This technique has then been formalised in the mid-1970s~\cite{tms76}. Discrete event modelling and simulation seem very close to point process models (dealing with events, directed graphs, continuous time, etc.). Against all expectations, as far as we know, there is no direct use of any discrete event simulation algorithm for point processes. Maybe, the sophistication of these algorithms being of the same order than the mathematical technicality of point processes, prevented any direct application. Besides, the continuous nature of the conditional intensity  associated to a point process with respect to the discreteness of event-based simulations could make appear the two domains as separated whereas discrete event theory is a computational specification of mathematical (continuous) systems theory~\cite{mesarovic} integrating more and more formally stochastic simulation concepts~\cite{tms2}. We hope to show here that both domains can take advantage from each other, by introducing new discrete-events models that act as generators of point processes. Especially, whereas discrete event simulation algorithms have been developed considering independently the components (nodes) of a system, a new algorithm for activity tracking simulation~\cite{cise-muzy} have been proposed to track activity (events from active nodes to children). The activity tracking algorithm is used here and proved to be the right tool for both simplifying usual discrete event algorithms (which are difficult to relate to usual point process algorithms) and efficiently simulating point processes.
        
        Our aim is to derive a new simulation algorithm, which generalises the algorithm of~\cite{doucet} to general multivariate point processes that are not necessarily Markovian, by exploiting the underlying network between the types, which is here a local independence graph. In Section \ref{setup}, the main mathematical background and notations are provided and the classical multivariate algorithm due to Ogata \cite{ogata} is explained. A simplified version in  discrete event terms and called \emph{full scan algorithm} is proposed. In Section \ref{sec:sched}, discrete event data structure and operations specific to point processes are designed. In Section \ref{dist}, after recalling the notion of  local independence graph \cite{didelez}, a new \emph{local graph algorithm} is presented. In Section \ref{sec:Hawkes}, we evaluate the computational complexities of both algorithms on Hawkes processes with piecewise constant interactions, which model easily neuronal spike trains~\cite{IEEE}. We show that in this case, for sparse graphs, new local graph algorithm clearly outperforms the classical Ogata's algorithm in its discrete event version.\\


	\section{Set-up}\label{setup}
    	\subsection{Mathematical framework}
        	A (univariate) point process $N$ in $\mathbb{R}_+$ is a random countable set of points of $\mathbb{R}_+$. For any subset $A$ of $\mathbb{R}_+$, $N(A)$ is the number of points that lie in $A$. 
        	
        	As real random variables might be defined by their density with respect to  Lebesgue measure, if it exists, a point process is characterised by its conditional intensity with respect to a given filtration or history $(\mathcal{F}_t,t\geq 0)$. For the mathematical details, we refer the reader to \cite{bremaud1981point}. Informally, the filtration or history at time $t-$, $\mathcal{F}_{t-}$, contains all the information that is needed to simulate the next point of $N$, when one is just before time $t$. It usually includes as generators, all the points $T\in N$ such that $T<t$ in particular. The (conditional) intensity of the point process $N$ is then informally defined ~\cite{bremaud1981point} by
        	\[
        	\lambda(t)=\lim_{dt\to 0} \frac{1}{dt}\mathbb{P}\left(\mbox{there is a point of } N \mbox{ in } [t,t+dt] \Big| \mathcal{F}_{t-}\right),
        	\]
        	for infinitesimal $dt$, where $\mathbb{P}\left(\mbox{there is a point of } N \mbox{ in } [t,t+dt] \Big| \mathcal{F}_{t-}\right)$ is the probability that a point appears in the interval $[t,t+dt]$ given what happened strictly before $t$ in the history. This  is a random process which, at time $t$, may depend in particular on all the past occurrences of the process itself, that is the $T<t$.
        	
        	A multivariate point process can be seen as a collection of $M$ different point processes $N_j$. With the \emph{time asynchrony hypothesis,} one can also consider equivalently the univariate joint point process $N=N_1\cup ... \cup N_M$ and say that for each point $t$ of $N$ there is one and only one subprocess $j$ such that $t\in N_j$. This $j$ is then called the mark of point $t$, or the node associated to $t$. 
        	
        	We are given  the set of intensities of each of the $N_j$, $t\mapsto\lambda_j(t)$, with respect to a common filtration $(\mathcal{F}_t,t\geq 0)$. Note that $\mathcal{F}_{t-}$  includes as generators,  all the points $T\in N$, the joint process such that $T<t$  as well as their respective marks.
        	
        	\paragraph{Examples}
        	Let us give just few basic examples
        	\begin{itemize}
        	    \item \textbf{Homogeneous Poisson processes} with rates $(\nu_i)_{i=1,...,M}$. In this case, all $\lambda_i$ are constant and not even random and for all $i$, $$\lambda_i(t)=\nu_i.$$ We see in this expression, that the intensities do not depend neither on time, nor on the previous occurrences. This is why one often refers to such dynamics as ``memoryless''. Since these processes do not interact, one can of course simulate each $N_i$ in parallel if need be. In this case, for each of them, it is sufficient to simulate the time elapsed until the next point, by an exponential variable of parameter $\nu_i$, independently from anything else. To unify frameworks, this exponential variable might also be seen as $-\log(U)/\nu_i$, with $U$ a uniform variable on $[0,1]$.
        	    \item \textbf{Inhomogeneous Poisson processes}  with time-dependent rates $(f_i)_{i=1,...,M}$. In this case, the $\lambda_i$'s are not necessarily constant and but they are still non random and for all $i$, $$\lambda_i(t)=f_i(t).$$
        	    Once again parallelization is possible, and for each individual process $N_i$ and given point $t_k^i$, one finds the next point $t_{k+1}^i$ by solving
        	    $$\int_{t_k^i}^{t_{k+1}^{i}}f_j(s)\dd s = -\log(U)$$
        	    \item \textbf{Linear multivariate Hawkes process} with spontaneous parameter $(\nu_j)_{j=1,...,M}$ and non negative interaction functions $(h_{j\to i})_{i,j=1,...,M}$ on $\mathbb{R}_+$. This process has intensity
        	    \begin{equation}\label{eq:hawkes}\lambda_i(t)=\nu_i + \sum_{j=1}^M \sum_{T \in N_j, T<t} h_{j\to i}(t-T).
        	    \end{equation}
        	    This process is used for many excitatory systems, especially the ones modelling the spiking activity of neurons \cite{IEEE}. It can be interpreted in this sense, informally: to a homogeneous Poisson process of rate $\nu_i$, which models the spontaneous activity of the neuron $i$, one adds extra-points coming from the interactions. Typically a point $T$ of mark (neuron) $j$ adds a term $h_{j\to i}(\delta)$, after  delay $\delta$ to the intensity of $N_i$ making the apparition of a new point at time $t=T+\delta$ more likely. In this sense there is an excitation of $j$ on $i$. 
        	    Here we see a prototypical example of global dependence between the marks. Each new point for each mark depends on all the points that have appeared before, with all the possible marks, preventing a brute force parallelization of the simulation. Except when the $h_{j\to i}$'s are exponentially decreasing~\cite{dassios}, this process is clearly not Markovian.  It is for this kind of general process that one needs efficient simulation algorithms.
        	\end{itemize}
    	
    	\subsection{Simulation of univariate processes}
            The time-rescaling theorem (see \cite{brown} or \cite{bremaud1981point} for more mathematical insight) states that if a point process $N$ has a conditional intensity $\lambda(t)$, and if  $$\forall t, \Lambda(t)=\int_0^t\lambda(s)\dd s,$$
            then $\mathcal{N}=\{\Lambda(T), T\in N\}$  is a Poisson process of rate 1. This is why, even for general point processes, it is always possible to find, by iteration  the next point of  $N$ by solving recursively, for all $k\in \mathbb{N}^*$ the set of positive natural numbers,
            \begin{equation}\label{eqrefinv}
            \int_{t_k}^{t_{k+1}}\lambda(s)\dd s = -\log(U)
            \end{equation}
            initializing the method with $t_0=0$.
            
            Of course, to be able to mathematically solve this easily, one needs to be able at time $t_k$ to compute $\lambda(t)$ on $(t_k,+\infty)$ if no other point occurs. This in particular happens if the filtration $\mathcal{F}_t$ is reduced to the filtration generated by the points themselves and this is what we will assume here. Of course all algorithms discussed here can easily be adapted to richer filtrations, as long as the computation of $\lambda(t)$ on $(t_k,+\infty)$ can be carried out (if no other point occurs).
            \\
            
            \noindent In this situation, two cases might happen, each of them leading to a different algorithm:
            \begin{description}
                \item[Transformation method:] The function $\lambda(t)$ on $(t_k,+\infty)$ (and if no other point occurs) has an easily computable primitive function with inverse $\Lambda^{-1}(t)$. Then \eqref{eqrefinv} reduces to 
                $$t_{k+1}=\Lambda^{-1}(-\log(U)+\Lambda(t_k)).$$
                \item[Thinning method:] It applies if the previous computation is not possible or easy  but one can still compute $\lambda^*(t)\geq \lambda(t)$ such that   $\lambda^*(t)$ has all the desired properties of the transformation method (typically $\lambda^*(t)$ is constant, with constant that might depend on the $t_\ell$ for $\ell\leq k$). Then the algorithm does as follows to compute a possible next point (cf. Algorithm \ref{alg:thinning}). If thinning for Poisson processes is due to \cite{lewis}, it has been generalized to general processes by Ogata~\cite{ogata}. One can find a complete proof in \cite{maria}.
                
            \end{description}

            \begin{algorithm}[H]
                    \begin{algorithmic}[1]
                        \caption{Thinning algorithm}\label{alg:thinning}
                        \State initialize $t_0^*\gets t_k$
                        \Repeat
            		      \State Generate next point $t^*$ after $t_0^*$ of a point process with intensity function $\lambda^*$ by the Transformation method.
            			  \State Generate $U\sim\mathcal{U}[0,1]$
            			    \If{$U > \lambda(t^*)/\lambda^*(t^*)$} \# Rejection
            					\State $t_0^*\gets t^*$
            				\EndIf
            			\Until $U \le \lambda(t^*)/\lambda^*(t^*)$
            			\State \Return $t_{k+1}\gets t^*$
            		\end{algorithmic}
            	\end{algorithm}
        
        \subsection{Discrete event version of classical multivariate algorithm for point processes}
            To simulate multivariate processes, Ogata \cite{ogata}  made an hybridation between two different notions : the thinning algorithm presented above and the attribution of marks. Indeed, since all processes $N_j$ are communicating with each other, Ogata's idea for the attribution of marks is to generate the next point of the aggregated process $N$ and, then, to decide for its mark. 
           
\begin{figure}[H]
        	 	\centering
        		\includegraphics[width=\textwidth]{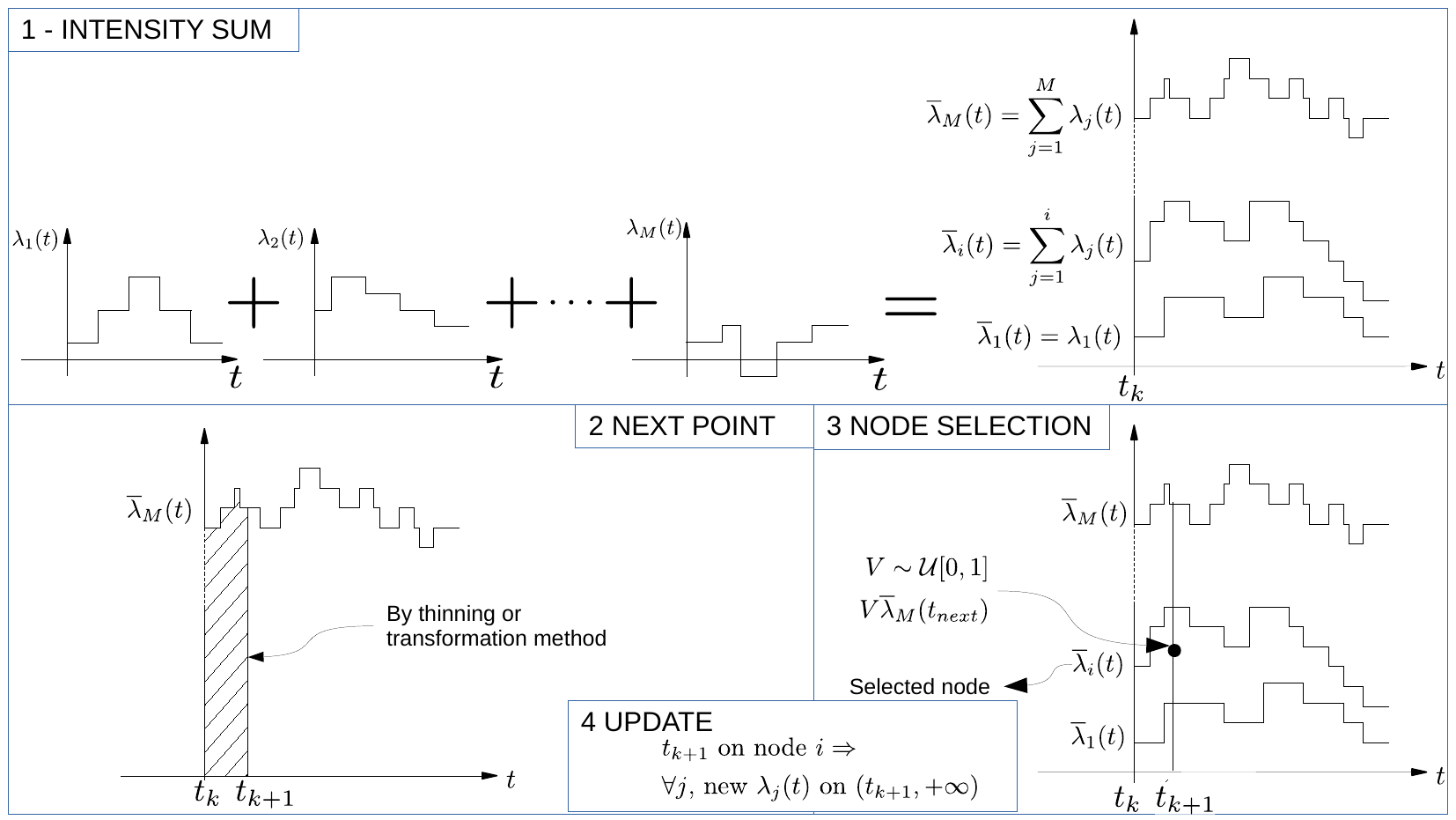}
        		\caption{Steps of the full scan algorithm for point processes. The intensities are piecewise constant (cf. Section~\ref{sec:Hawkes}).}\label{fig:centra}
        	\end{figure}

           In the present article, we choose to discard the thinning part, which can be added to all the algorithms that we derive here. The  attribution part of Ogata's multivariate algorithm \cite{ogata}, is referred in the sequel as \emph{full scan}, 
           because the intensities of all nodes of the graph need to be scanned and updated at each time stamp $t_k$. Once the next point of $N$ is decided,  the basic idea for the attribution of marks is to attribute them at random, the distribution taking into account the relative value of the intensity of each subprocess.  The main steps of this  algorithm are presented in Figure \ref{fig:centra} for a visual representation of the method. More details about the algorithm steps are provided through the Hawkes application in Section~\ref{sec:Hawkes}.

        	\begin{algorithm}[H]
        		\begin{algorithmic}[1]
        			\State $t_0 \leftarrow 0$
        			\While{$t_k < T$}
            			\State \textbf{Compute intensity sums} $\sum_{j=1}^i \lambda_j(t) = \overline{\lambda}_i(t)$, for $i \in \{1,...,M \}$ on $t\in(t_k,+\infty)$\label{algline:cent-comp-sum}
            			\State \textbf{Get} by simulation  $t_{k+1}$ as the \textbf{next point} of a univariate point process of intensity $\overline{\lambda}_M(t)$\label{alg-line:cent-next-point}
            			\State \textbf{Select the unique possible node $i_{k+1}$}, such that  $\frac{\overline{\lambda}_{i_{k+1}-1}(t_{k+1})}{\overline{\lambda}_M(t_{k+1})} < V \leq \frac{\overline{\lambda}_{i_{k+1}}(t_{k+1})}{\overline{\lambda}_M(t_{k+1})}$, with $V\sim \mathcal{U}[0,1]$ and $\overline{\lambda}_0=0$
            			\State \textbf{Update intensities } $\lambda_j$ on  $(t_{k+1},+\infty), \forall j \in \{1,...,M \}$ \label{algline:cent-update}
            			\State $k \leftarrow k+1$
                	\EndWhile
        		    \State \textbf{return} points $(t_1, ..., t_{k-1})$ and associated nodes $(i_1, ..., i_{k-1})$
        		\end{algorithmic}
        		\caption{Full scan multivariate algorithm modified from \cite{ogata}}\label{alg:cent}
        		\AddNoteOneLine[black]{3}{3}{Step a/}
        		\AddNoteOneLine[orange]{4}{4}{Step b/}
        		\AddNoteOneLine[purple]{5}{5}{Step c/}
        		\AddNoteOneLine[green]{6}{6}{Step d/}
        	\end{algorithm}
        


        	Original Ogata's algorithm uses thinning at step~\ref{alg-line:cent-next-point} of Algorithm~\ref{alg:cent}. However the complexity of a thinning step is difficult to evaluate because it depends on both the complexity of the upper-bounding function $\lambda^*$ and how far this function is from $\lambda$, which influences how much time the thinning algorithm rejects. Therefore for a clear evaluation of the complexity, we focused on simulations where the transformation method is doable, typically when the intensities are piecewise constant.
        	
        		\section{Specific discrete event data structures and operations}\label{sec:sched}
        		Before introducing our new algorithm, we present a particular structure, which is very important for discrete events algorithm : the scheduler.
        		
        		A \emph{scheduler} $Q$ is a data structure, which can be represented as an ordered set of \emph{events}, and which is provided with a set of operations for ensuring the correct order of the events. The events are noted $ev_i=(t_i,v_i)$, where $t_i$ is the \emph{event time} and $v_i$ is the \emph{event value}. The events in the scheduler are increasingly ordered in time, i.e., $ev_i,ev_{j} \in Q$, $ev_i < ev_j \iff t_i < t_j$. The length of the scheduler is noted $|Q|$. 
        		
        		\begin{figure}[H]
		            \centering \includegraphics[width=0.6\textwidth]{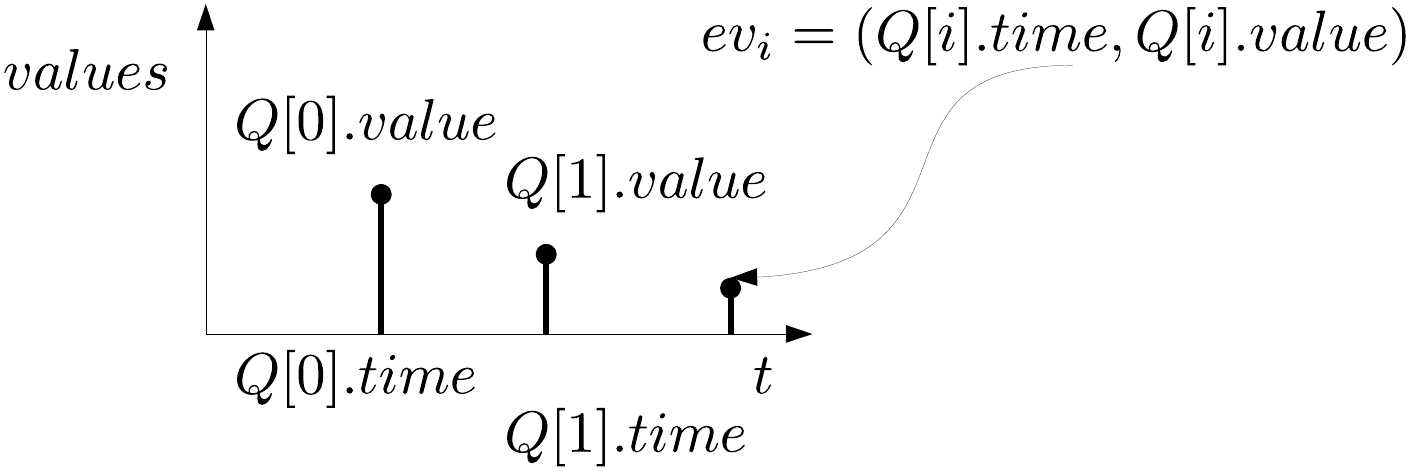}
		            \caption{Example of events in scheduler $Q$. For an event $ev_i$ in the scheduler, the \emph{value} is accessed by $Q[i].value$ and the \emph{time} is accessed by $Q[i].time$, with $i=0$ the first event index.}
		            \label{fig:eventexample}
		        \end{figure}

    		    In both full scan and local graph algorithms schedulers are used. They are usually implemented using one of the many kinds of self-balancing binary tree. The choice of structure (AVL, red-black, etc.) depends on corresponding operation complexity. Here we choose the red-black self balancing tree, which exhibits the best performance with respect to other usual self-balancing trees \cite{heger}. 
    		    
    		    To make a scheduler, the red-black tree is equipped with a set of classical (\textbf{insert}, \textbf{remove}, \textbf{upper bound} and \textbf{lower bound}) and non-classical operations. All these operations, except stated otherwise, have time complexity bounded by the logarithm $\log_2$ of the number of elements in the set. In addition it is supposed that any element of the scheduler can be accessed with a constant time. This may not be the case depending on the language used for the implementation. However, it is true for C++ language, which was used in our implementation.

            \subsection{Basic \emph{scheduler operations}}
            
            We describe and illustrate the set of operations completing the red-black tree to form the scheduler data structure needed for our algorithms. Any operation on a scheduler $Q$ directly modifies it. Also, to simplify the operations, it is supposed that all events stored in a scheduler $Q$ are unique.
            
            \begin{description}
            \item[Access element operation] $(t^i,v^i)_{i\in\{0,\dots,|Q|-1\}}$ is the list of events stored in scheduler $Q$ and sorted by ascending values of $t$. Then operation $Q[i]$ returns event $(t^i,v^i)$ with constant time complexity $\mathscr{O}(1)$.
            
            \item[Insert operation] of an event $(t,v)\in\mathbb{R}_+\times\mathbb{R}$ (cf. Figure \ref{fig:insert}): $Q \oplus (t,v)$ inserts the event in the tree. This operation has a total maximum complexity of $\mathcal{O}(log_2(|Q|))$, to find the place of the event and rebalance the tree.
        		    
        		    \begin{minipage}{\linewidth}
        		        \centering
        		        \begin{tikzpicture}
        		            \draw[-latex] (-0.1, 0) -- (3.5, 0) node[below right] {time};
        		            \draw[-latex] (0, -0.1) -- (0, 2) node[left] {values};
        		            
        		            \draw[-o, dotted] (0.1, 0) -- (0.1, 1);
        		            \draw[-o, dotted] (1, 0) node [below] {$t_1$} -- (1, 1);
        		            \draw[-o, dotted] (2, 0) node [below] {$t_2$} -- (2, 1.5);
        		            \draw[-o, dotted] (3, 0) node [below] {$t$} -- (3, 0.5) node [left] {$v$};
        		            
        		            \draw[-*, dashed] (5.5, 0) node [below, align=center] {$t_1<t_*<t_2$\\or $t=t_*$} -- (5.5, 1.75) node [left] {$v_*$};
        		            
        		            \node at (4, 0.75) {\scalebox{3}{${\bm\oplus}$}};
        		            \node at (6.5, 0.75) {\scalebox{3}{$\bm=$}};
        		            
        		            \draw[-latex] (7.4, 0) -- (11, 0) node[below right] {time};
        		            \draw[-latex] (7.5, -0.1) -- (7.5, 2) node[left] {values};
        		            
        		            \draw[-o, dotted] (7.6, 0) -- (7.6, 1);
        		            \draw[-o, dotted] (8.5, 0) node [below] {$t_1$} -- (8.5, 1);
        		            \draw[-*, dashed] (9, 0) node [below] {$<t_*<$} -- (9, 1.75);
        		            \draw[-o, dotted] (9.5, 0) node [below] {$t_2$} -- (9.5, 1.5);
        		            \draw[-*] (10.5, 0) node [below] {$t=t_*$} -- (10.5, 2.25) node [left] {$v+v_*$};
        		        \end{tikzpicture}
        		        
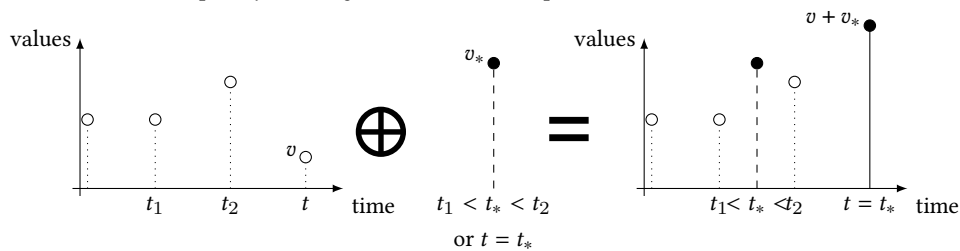
\captionof{figure}{A graphical example of the insertion of a new event $(t_*, v_*)$  inside a scheduler. There are two cases: if the event time is unmatched in the set of event times already present in the scheduler (case $t_1<t_*<t_2$), the event is just inserted in the right place; otherwise (case $t=t_*$) the event in the scheduler with the same time  has its value increased by the value $v_*$ of the new event.}
        		        \label{fig:insert}
        		    \end{minipage}
        		    
        		   \item[Remove operation] of an event (cf. Figure \ref{fig:remove}):  $Q \ominus (t,v)$, removes the event $(t,v)$ from the tree. This operation has a maximum complexity of $\mathscr{O}\left(\log_2(|Q|)\right)$, to find the place of the event and rebalance the tree. If the index $i$ of the element is known, the remove operation can be executed in constant time.
        		   
        		     \begin{minipage}{\linewidth}
        		        \centering
        		        \begin{tikzpicture}
        		            \draw[-latex] (-0.1, 0) -- (3.5, 0) node[below right] {time};
        		            \draw[-latex] (0, -0.1) -- (0, 2) node[left] {values};
        		            
        		            \draw[-o, dotted] (0.1, 0) -- (0.1, 1);
        		            \draw[-o, dotted] (1, 0) node [below] {$t_1$} -- (1, 1);
        		            \draw[-o, dotted] (2, 0) node [below] {$t_2$} -- (2, 1.5);
        		            \draw[-o, dotted] (3, 0) node [below] {$t$} -- (3, 0.5);
        		            
        		            \node at (4.5, 1) {\scalebox{3}{${\bm\ominus (t,v)}$}};
        		            \node at (6.5, 1) {\scalebox{3}{$\bm=$}};
        		            
        		            \draw[-latex] (7.4, 0) -- (11, 0) node[below right] {time};
        		            \draw[-latex] (7.5, -0.1) -- (7.5, 2) node[left] {values};
        		            
        		            \draw[-o, dotted] (7.6, 0) -- (7.6, 1);
        		            \draw[-o, dotted] (8.5, 0) node [below] {$t_1$} -- (8.5, 1);
        		            \draw[-o, dotted] (9.5, 0) node [below] {$t_2$} -- (9.5, 1.5);
        		        \end{tikzpicture}
        		        
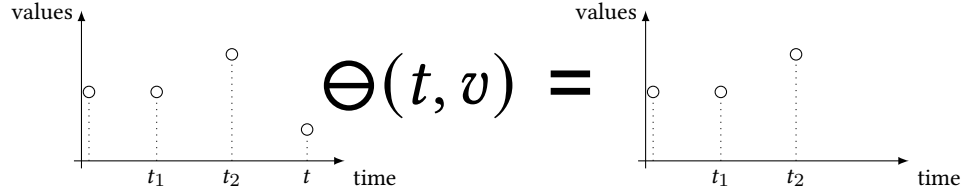
\captionof{figure}{A graphical example of the removal of an event given its time $t$.}
        		        \label{fig:remove}
        		    \end{minipage}
                
                \item [Remove first operation] $Q^*$ over the scheduler $Q$, which removes the first event $Q[0]=(t^0,v^0)$ from the scheduler, the second event becoming the first. Operation $\cdot^*$ has complexity $\mathcal{O}(1)$.
                
                \item [Prune operation] of the scheduler $Q^t$ (cf. Figure \ref{fig:prune}) removes all events $(t^*,v^*)$ from $|Q|$ whose time value $t^*\leq t$. The operation $Q^t$ has complexity $\mathcal{O}\left(log_2(|Q|)+|\textrm{number of events until time }t|\right)$.
                
                  \begin{minipage}{\textwidth}
        		        \centering
        		        \begin{tikzpicture}
        		            \draw[-latex] (-0.1, 0) -- (3.5, 0) node[below right] {time};
        		            \draw[-latex] (0, -0.1) -- (0, 2) node[left] {values};
        		            \node[below] at (1.5, 0) {$t$};
        		            
        		            \draw[-o] (0.1, 0) -- (0.1, 1);
        		            \draw[-o] (1, 0) -- (1, 1);
        		            \draw[-o] (2, 0) -- (2, 1.5);
        		            \draw[-o] (3, 0) -- (3, 0.5);
        		            
        		            \node at (4.5, 1) {\scalebox{3}{$\bm{Q^t}$}};
        		            \node at (6, 1) {\scalebox{3}{$\bm=$}};
        		            
        		            \draw[-latex] (6.9, 0) -- (10.5, 0) node[below right] {time};
        		            \draw[-latex] (7, -0.1) node [below] {$t$} -- (7, 2) node [left] {values};
        		            
        		            \draw[-o] (7.5, 0) -- (7.5, 1.5);
        		            \draw[-o] (8.5, 0) -- (8.5, 0.5);
        		        \end{tikzpicture}
        		        
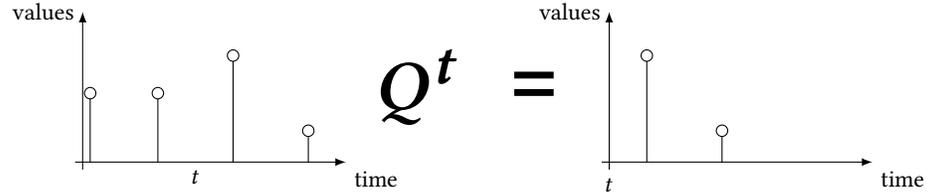
\captionof{figure}{A graphical example of the prune operation : Here all the points with a time less or equal to $t$ (here the first two points of this scheduler) are removed. The time complexity of the operation is $\mathscr{O}(\log_2(4)+2)$ (scheduler of size $|Q|=4$ and $2$ events removed).}
        		        \label{fig:prune}
        		    \end{minipage}
                
                 \item [Upper and lower bound operations] (cf. Figure \ref{fig:lower}): Upper bound operation: $\left\lceil t \right\rceil_Q$ and lower bound operation $\left\lfloor t \right\rfloor_Q$, return the smaller event $(t^*,v^*)$ verifying respectively $t^*>t$ and $t^*\geq t$. In both cases the time complexity of the operation is $\mathcal{O}(log_2(|Q|))$.
                 
        		    \begin{minipage}{\textwidth}
        		        \centering
        		        \begin{tikzpicture}
                		            \draw[-latex] (-0.1, 0) -- (3.5, 0) node [below right] {time};
                		            \draw[-latex] (0, -0.1) -- (0, 2) node[left] {values};
                		            
                		            \draw[-o] (0.1, 0) node[below] {$t_i$}  -- (0.1, 1);
                		            \draw[-o] (1, 0) node[below] {$t_j$} -- (1, 1);
                		            \draw[-o] (2, 0) node[below] {$t_k$} -- (2, 1.5);
                		            \draw[-o] (3, 0) node[below] {$t_\ell$} -- (3, 0.5);
        		          
        		            
        		            
        		        \end{tikzpicture}
        		        
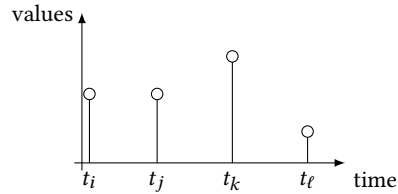
\captionof{figure}{Upper/lower bound operations on an example of scheduler $Q$. For $t_k<t<t_\ell$ the \emph{upper bound operation} consists of $\left\lceil t \right\rceil_Q = \ell $. The \emph{lower bound operation} consists of $\left\lfloor t \right\rfloor_Q = k $.}
        		        \label{fig:lower}
        		    \end{minipage}
                
                
            \end{description}

            \subsection{Using a scheduler to represent piecewise constant functions}
\begin{figure}
                \centering
                \begin{tabular}{c|c|c|}
                    \textsc{Function} & \textsc{Encoding} & \textsc{Summing}\\ \hline
                    \begin{tikzpicture}[baseline=0]
                        \draw[-latex] (0,0) -- (0,2) node [left] {$h,h^\prime$};
                        \draw[-latex] (0,0) node[below] {\scriptsize$t_k$} -- (3.75,0);
                        
                        \node at (1,0) [below] {\scriptsize$t_{k+1}$};
                        \node at (2.5,0) [below] {\scriptsize$t_{k+2}$};
                        \node at (1.75,0) [below] {\scriptsize$t_{k+3}$};
                        \node at (3,0) [below] {\scriptsize$t_{k+4}$};
                        
                        \path[draw] (0,0.5) -- node[above] {\tiny$H_0$} (1,0.5) -- (1,1.5) -- node[above] {\tiny$H_1$} (2.5,1.5) -- (2.5,0.75) -- node[above] {\tiny$H_2$} (3.5,0.75);
                        
                        \path[draw, dashed] (1.75,0) -- node[left] {$H_1^\prime$} (1.75,0.5) -- (3,0.5) -- node[right] {$H_2^\prime$} (3,0);
                    \end{tikzpicture} &
                    \begin{tikzpicture}[baseline=0]
                        \draw[-latex] (0,0) -- (0,2) node [left] {h};
                        \draw[-latex] (0,0) node[below] {\scriptsize$t_k$} -- (3.75,0);
                        
                        \node at (1,0) [below] {\scriptsize$t_{k+1}$};
                        \node at (2.5,0) [below] {\scriptsize$t_{k+2}$};
                        
                        \draw[ultra thick] (0,0) -- node[right] {\tiny$\bm H_0$} (0,0.5);
                        \draw[ultra thick] (1,0) -- node[right] {\tiny$\bm H_1$} (1,1.5);
                        \draw[ultra thick] (2.5,0) -- node[right] {\tiny$\bm H_2$} (2.5,0.75);
                        
                        \path[draw] (0,0) -- (0,0.5) -- (1,0.5) -- (1,1.5) -- (2.5,1.5) -- (2.5,0.75) -- (3.5,0.75);
                    \end{tikzpicture} &
                    \begin{tikzpicture}[baseline=0]
                        \draw[-latex] (0,0) -- (0,2) node [left] {$h+h^\prime$};
                        \draw[-latex] (0,0) node[below] {\scriptsize$t_k$} -- (3.75,0);
                        
                        \node at (1,0) [below] {\scriptsize$t_{k+1}$};
                        \node at (2.5,0) [below] {\scriptsize$t_{k+2}$};
                        \node at (1.75,0) [below] {\scriptsize$t_{k+3}$};
                        \node at (3,0) [below] {\scriptsize$t_{k+4}$};
                        
                        \draw[ultra thick] (0,0) -- node[right] {\tiny$\bm H_0$} (0,0.5);
                        \draw[ultra thick] (1,0) -- node[right] {\tiny$\bm H_1$} (1,1.5);
                        \draw[ultra thick] (1.75,0) -- node[right] {\tiny$(*)$} (1.75,2);
                        \draw[ultra thick] (2.5,0) -- node[right] {\tiny$(**)$} (2.5,1.25);
                        \draw[ultra thick] (3,0) -- node[right] {\tiny$\bm H_2$} (3,0.75);
                        
                        \path[draw] (0,0) -- (0,0.5) -- (1,0.5) -- (1,1.5) -- (1.75,1.5) -- (1.75,2) -- (2.5,2) -- (2.5,1.25) -- (3,1.25) -- (3,0.75) -- (3.5,0.75);
                        
                        \node at (3.5,2) {\tiny$(*): \bm{H_1+H_1^\prime}$};
                        \node at (3.5,1.75) {\tiny$(**): \bm{H_2+H_2^\prime}$};
                    \end{tikzpicture}\\ \hline
                    
                    \begin{tikzpicture}[baseline=0]
                        \draw[-latex] (0,0) -- (0,2) node [left] {$h,h^\prime$};
                        \draw[-latex] (0,0) node[below] {\scriptsize$t_k$} -- (3.75,0);
                        
                        \node at (1,0) [below] {\scriptsize$t_{k+1}$};
                        \node at (2.5,0) [below] {\scriptsize$t_{k+2}$};
                        \node at (1.75,0) [below] {\scriptsize$t_{k+3}$};
                        \node at (3,0) [below] {\scriptsize$t_{k+4}$};
                        
                        \path[draw] (0,0) -- node[right] {\tiny$\Delta H_0$} (0,0.5) -- (1,0.5) -- node[right] {\tiny$\Delta H_1$} (1,1.5) -- (2.5,1.5) -- node[right] {\tiny$\Delta H_2$} (2.5,0.75) -- (3.5,0.75);
                        
                        \path[draw, dashed] (1.75,0) -- node[left] {\tiny$\Delta H^\prime_0$} (1.75,0.5) -- (3,0.5) -- node[right] {\tiny$\Delta H^\prime_1$} (3,0);
                    \end{tikzpicture} &
                    \begin{tikzpicture}[baseline=0]
                        \draw[-latex] (0,0) -- (0,2) node [left] {h};
                        \draw[-latex] (0,0) node[below] {\scriptsize$t_k$} -- (3.75,0);
                        
                        \node at (1,0) [below] {\scriptsize$t_{k+1}$};
                        \node at (2.5,0) [below] {\scriptsize$t_{k+2}$};
                        
                        \draw[ultra thick] (0,0) -- node[right] {\tiny$\bm{\Delta H_0}$} (0,0.5);
                        \draw[ultra thick] (1,0.5) -- node[right] {\tiny$\bm{\Delta H_1}$} (1,1.5);
                        \draw[ultra thick] (2.5,1.5) -- node[right] {\tiny$\bm{\Delta H_2}$} (2.5,0.75);
                        
                        \path[draw] (0,0) -- (0,0.5) -- (1,0.5) -- (1,1.5) -- (2.5,1.5) -- (2.5,0.75) -- (3.5,0.75);
                    \end{tikzpicture} &
                    \begin{tikzpicture}[baseline=0]
                        \draw[-latex] (0,0) -- (0,2) node [left] {$h+h^\prime$};
                        \draw[-latex] (0,0) node[below] {\scriptsize$t_k$} -- (3.75,0);
                        
                        \node at (1,0) [below] {\scriptsize$t_{k+1}$};
                        \node at (2.5,0) [below] {\scriptsize$t_{k+2}$};
                        \node at (1.75,0) [below] {\scriptsize$t_{k+3}$};
                        \node at (3,0) [below] {\scriptsize$t_{k+4}$};
                        
                        \draw[ultra thick] (0,0) -- node[right] {\tiny$\bm{\Delta H_0}$} (0,0.5);
                        \draw[ultra thick] (1,0.5) -- node[right] {\tiny$\bm{\Delta H_1}$} (1,1.5);
                        \draw[ultra thick] (1.75,1.5) -- node[left] {\tiny$\bm{\Delta H^\prime_0}$} (1.75,2);
                        \draw[ultra thick] (2.5,2) -- node[right] {\tiny$\bm{\Delta H_2}$} (2.5,1.25);
                        \draw[ultra thick] (3,1.25) -- node[right] {\tiny$\bm{\Delta H^\prime_1}$} (3,0.75);
                        
                        \path[draw] (0,0) -- (0,0.5) -- (1,0.5) -- (1,1.5) -- (1.75,1.5) -- (1.75,2) -- (2.5,2) -- (2.5,1.25) -- (3,1.25) -- (3,0.75) -- (3.5,0.75);
                    \end{tikzpicture}
                \end{tabular}
                \caption{Example of a straightforward encoding of a piecewise constant function (first line) and an optimised one (second line). We use the notation $H_n=h(t_{k+n}), n\in\mathbb{N}$ for concision. The \textsc{Function} column represents the piecewise constant function $h$ to encode (plain lines), as well as another function $h^\prime$ (dashed lines, see column \textsc{Summing}). Column \textsc{Encoding} represents in bold the events $(t_n,h(t_n))$ encoded by a scheduler representing the function $h$. Column \textsc{Summing} represents in bold the values encoded by a scheduler now representing the function $h+h^\prime$.}\label{fig:breakpoint_tab}
            \end{figure}
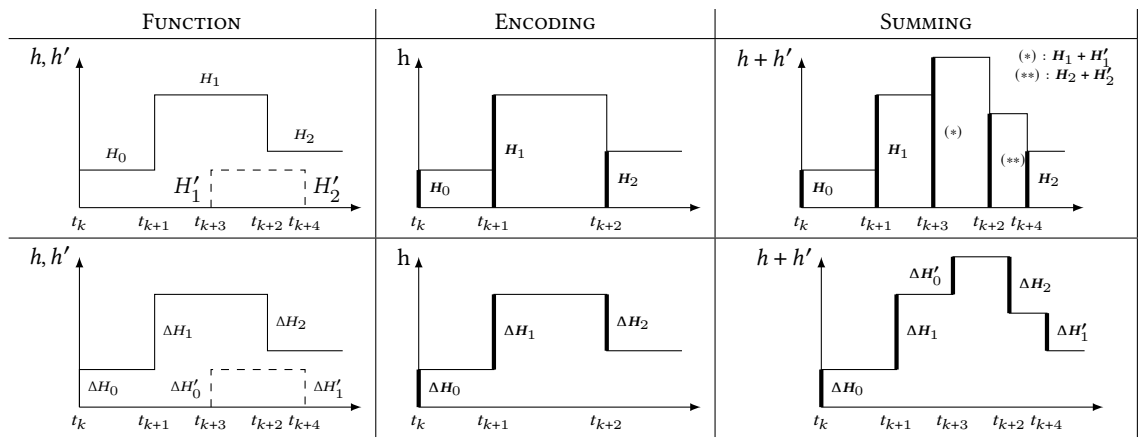

            Let $h\colon\mathbb{R}^*\to\mathbb{R}, t\mapsto h(t)$ be a piecewise constant function with finite support $S\subseteq\mathbb{R}^*$. A scheduler can encode piecewise constant functions on $S$. There are two possible methods to achieve this:
            \begin{enumerate}
                \item Let each discontinuity $(t,h(t))$ be its own event in the scheduler $Q$ representing the function $h$ (see top part of Figure~\ref{fig:breakpoint_tab}).
                \item Represent the discontinuity not as a new value $(t,h(t))$ but as an increment, that is a difference between the new value of $h$ after the discontinuity and the value of $h$ just before: $(t,h(t)-h(t_-))$ (see bottom part of Figure~\ref{fig:breakpoint_tab}). 
            \end{enumerate}
            
            The first representation is more straightforward, and is very efficient when the function is stored only for lookups: the function can then be stored in an array and a dichotomic search algorithm be used to locate any event in logarithmic time.
            
            However, if discontinuities must be introduced or removed, for instance by adding another piecewise constant function to $h$, then some of the events stored in the scheduler may need to be changed or moved in the scheduler, thus creating an undesirable overhead. For instance in the first line of figure~\ref{fig:breakpoint_tab}, a piecewise function $h$ (first column) is represented and encoded (second column) with the straightforward method by a scheduler. Last column shows the new encoding representing a sum. The discontinuity at $t_{k+2}$ had to be modified because of the introduction of discontinuities at $t_{k+3}$ and $t_{k+4}$. For a function with many more discontinuities, or when summing two piecewise functions of similar size, the operation would necessitate to modify a large part of the already represented points.
            
            The solution we proposed is to store not the values $h(t)$ of the function $h$ at a time $t$ where a discontinuity happens, but the variation $\Delta h_t=h(t)-h(t_-)$ of the value of $h$ during the discontinuity. An example is represented on the second line of Figure~\ref{fig:breakpoint_tab}. 
            
\begin{figure}
        		             \begin{tikzpicture}
        		            \draw[-latex] (-0.1, 0) -- (4.5, 0);
        		            \draw[-latex] (0, -0.1) -- (0, 2) node[left] {$\sched_i$};
        		            
        		            \draw[-o, dotted] (0.1, 0) -- (0.1, 1);
        		            \draw[-o, dotted] (1, 0) -- (1, 1);
        		            \draw[-o, dotted] (2, 0) -- (2, 1.5);
        		            \draw[-o, dotted] (3, 0) -- (3, 0.5);
        		            
        		            \draw[-latex] (-0.1, -5) -- (4.5, -5);
        		            \draw[-latex] (0, -5.1) -- (0, -2.5) node[left] {$\sched_j$};
        		            
        		            \draw[-o, dashed] (0.5, -5) -- (0.5, -4);
        		            \draw[-o, dashed] (1, -5) -- (1, -3.5);
        		            \draw[-o, dashed] (2, -5) -- (2, -4.5);
        		            \draw[-o, dashed] (4, -5) -- (4, -4);
        		            
        		            \node at (2.5, -1.5) {\scalebox{3}{$\bm\cup$}};
        		            \node at (5.5, -1.5) {\scalebox{3}{$\bm=$}};
        		            
        		            \draw[-latex] (6.9, -2.5) -- (12, -2.5);
        		            \draw[-latex] (7, -2.6) -- (7, 0) node [left] {$\sched_k$};
        		            
        		            \draw[-o, dotted] (7.1, -2.5) -- (7.1, -1.5);
        		            \draw[-o, dashed] (7.5, -2.5) -- (7.5, -1.5);
        		            \draw[-o] (8, -2.5) -- (8, 0);
        		            \draw[-o] (9, -2.5) -- (9, -0.5);
        		            \draw[-o, dotted] (10, -2.5) -- (10, -2);
        		            \draw[-o, dashed] (11, -2.5) -- (11, -1.5);
        		        \end{tikzpicture}
        		        \captionof{figure}{A graphical example of the union of two schedulers. The events of $\sched_i$ are represented as dotted, while the events of $\sched_j$ are dashed. In the merged scheduler $\sched_k$, the values of events at the same time in both schedulers $i$ and $j$ are summed and the resulting event is represented with a continuous line.}
        		        \label{fig:union}
        			\end{figure}
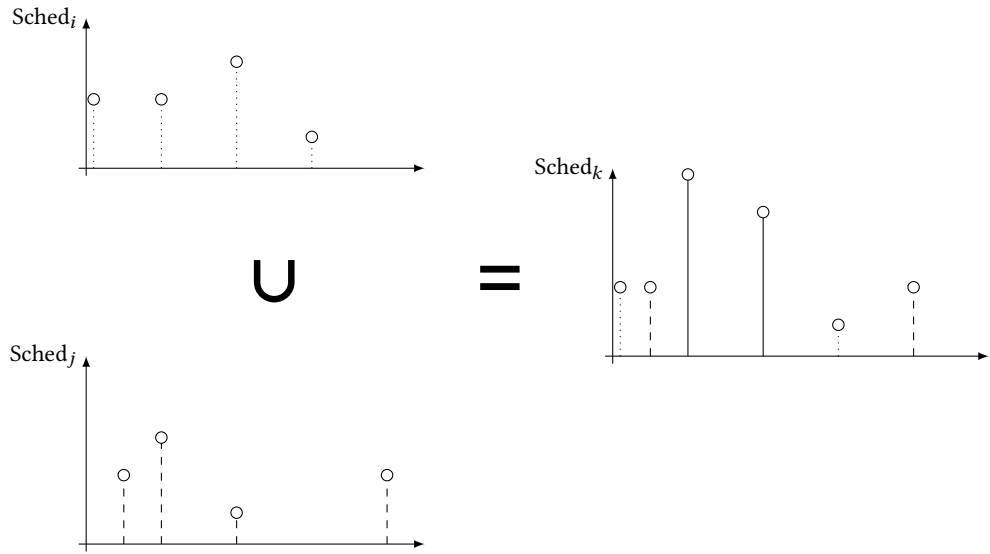

\begin{figure}
        			\begin{center} \includegraphics[width=0.6\textwidth]{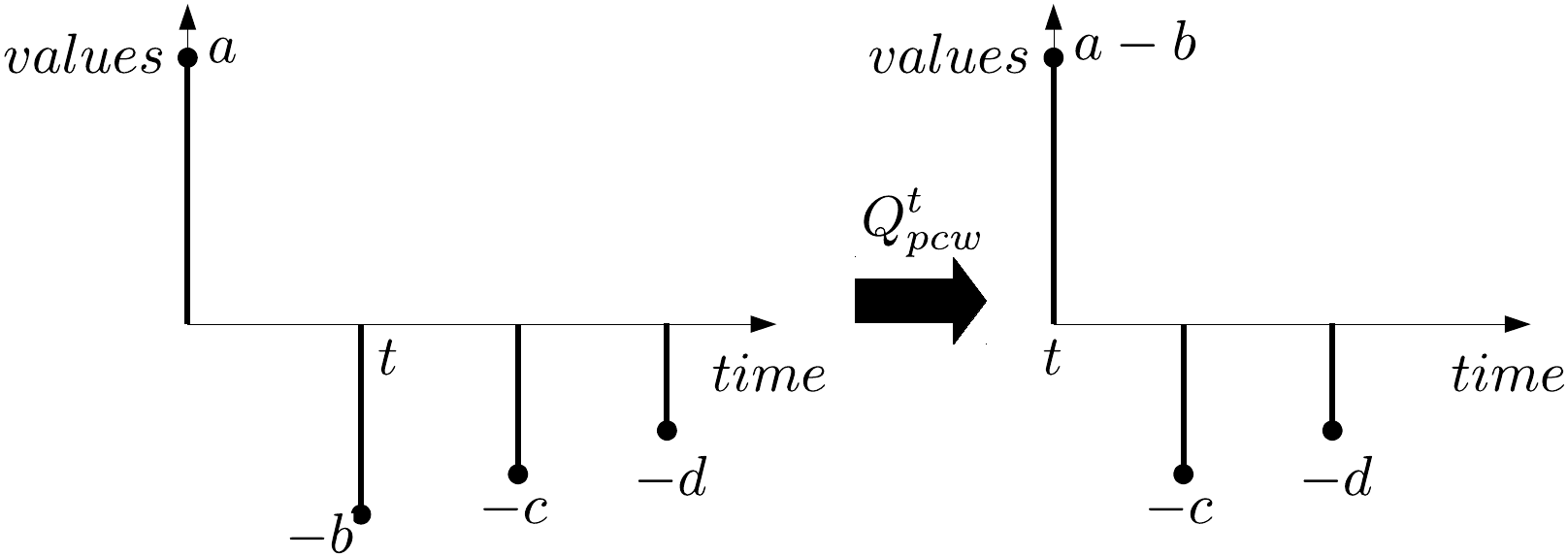}
                	\end{center}                 \captionof{figure}{Example of piecewise prune operation $Q^t_{pcw}$.\label{fig:pcw-shift}}
                    
                \end{figure}

\begin{figure}
                    \centering
                    \includegraphics[width=0.6\textwidth]{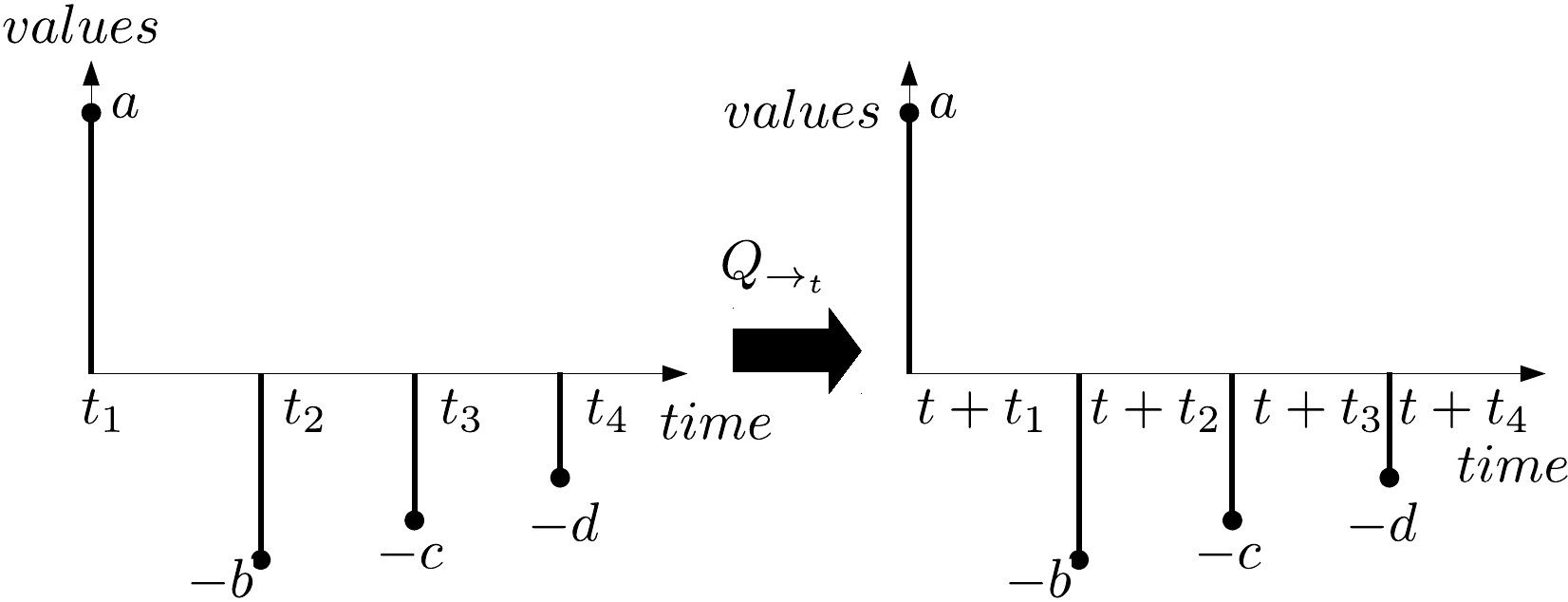}
                    \caption{The shift operation $Q_{\rightarrow_t}$}
                    \label{fig:shift}
                \end{figure}

            Let us now present two different operations for piecewise contant functions with this encoding: 

            \begin{description}
                 \item [Piecewise sum,] which is in fact a union operation over two schedulers $Q, Q'$ encoding two piecewise constant functions (cf. Figure \ref{fig:union}): The operation  $Q \cup Q'$ has complexity $\mathcal{O}(min(|Q|,|Q'|)log_2(max(|Q|,|Q'|)))$, since the smallest scheduler is inserted into the largest scheduler.

        		 \item [Piecewise prune operation] of the scheduler $Q^t_{pcw}$ (cf. Figure \ref{fig:pcw-shift}): removes from scheduler $Q$ all the events $(t^*,v^*)$ with $t^*\leq t$, while accumulating all the values $v^*$ up to time $t$. At the end of the operation the scheduler begins with an event of the form $(t, \sum v^*)$. The operation $Q^t_{pcw}$ has complexity $\mathcal{O}(log_2(|Q|)+|\textrm{number of events until time }t|)$.
        		 
        	    \item [Shift operation] of the scheduler $Q_{\rightarrow_t}$ (cf. Figure \ref{fig:shift}): shifts all points by $t$. The operation $Q_{\rightarrow_t}$ has complexity $\mathcal{O}(|Q|)$.        	    
        	\end{description}
        	
	\section{Local graph algorithm for point processes}\label{dist}
	    \subsection{Local independence graph}
\begin{figure}[H]
    			\begin{center}
    				\begin{tikzpicture}
    					\pic {graph=0};
    				\end{tikzpicture}
    			\end{center}
    			\caption{Example of local independence graph. With this graph, $ch(2)=\{3,6\}$ and $pa(2)=\{1,5\}$. As indicated by the difference of colour, a point with mark $2$ shall impact the point generation only for $\{2\}\cup\{3,6\}$.\label{fig:dist-ind1}}
		    \end{figure}
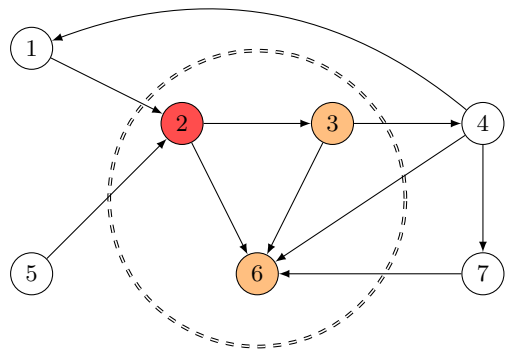

	        Local independence graphs are fully presented in a sound mathematical form in \cite{didelez}. For a given multivariate point process $N_j, {j=1,...,M}$, the corresponding local independence graph is a directed graph on the nodes $j=1,...,M$ (see for instance Figure \ref{fig:dist-ind1}). We assume for sake of simplicity that the filtration is reduced to the internal history, that is $\mathcal{F}_t$ is generated only by the $T<t$ in $N=N_1\cup...\cup N_M$ and their associated mark or node.
	
	        To explain more fully what a local independence graph means, we need to define rougher filtration. For a subset $I \subset \{1,...,M\}$, $\mathcal{F}_t^I$ is the filtration generated by  the $T<t$ in $\cup_{i\in I} N_i$ and their associated  node.
	
	        In a local independence graph, the absence of edge $j\to i$ means that the apparition of a point at time $t$ in $N_i$ is independent from $\mathcal{F}_{t-}^{\{j\}}$ conditionally to $\mathcal{F}_{t-}^{ \{j\}^c}$, where $ \{j\}^c= \{1,...,M\} \setminus \{j\}$.
	
	        So this means that for every time $t$,  the intensity  $\lambda_i(t)$ of $N_i$ 
	        does not depend directly on the positions of the points of $N_j$ strictly before $t$. 
	
	        This extends directly to the notion of parents and children in the graph. For a given node $i$, one defines \[
	            pa(i)=\{j, j\to i \mbox{ is in the graph} \} \mbox{ and } ch(i)=\{j, i\to j \mbox{ is in the graph}\}.
	        \]
	
            Therefore it means that the intensity $\lambda_i(t)$  at time $t$ of $N_i$ 
            in fact only depends on the points of $N_j$ for $j\in pa(i)$ strictly before $t$. 
	
            Conversely, a point on $N_i$ directly impacts the occurrence of points for $N_j$ for $j\in ch(i)$. Note that in any case, it also impacts the next point of $N_i$ because even for a Poisson process without memory one needs by the transformation method to know $t_k$ for finding $t_{k+1}$. However, it will not have any direct impact on the future points of $N_j$ for $j\not\in ch(i)\cup\{i\}$.

		For instance, for linear Hawkes process (cf. Equation~\ref{eq:hawkes}), $ch(i)=\{j / h_{i\to j}\neq 0\}.$
		
	    \subsection{Local-graph algorithm}
	        
	  
	       The children of each node are stored in a simple one dimensional array, whose indexes are the node indexes, and whose cells contain vectors of the children indexes. So accessing a node simply costs $\mathcal{O}(1)$.
	        
            
           Because of the interpretation of $I=ch(i)\cup\{i\}$ of a given node $i$ given above in the local independence graph, it means that in fact, after having simulated $t_k$ with mark/node  $i_k$ in the joint process, we know that only the next points of $N_j$ for $j\in I$ have to be modified. 
            
            At simulation level, discrete events are used to track activity nodes associated to selected points (time stamps) to their children. Discrete events are stored into a scheduler $Q$ of events $ev_i=(t_{next}^i,i)$, where $t_{next}^i$ is the \emph{possible next point} associated to node $i$.
        			
\begin{figure}
            	\begin{center}
                     \includegraphics[page=1,width=\textwidth]{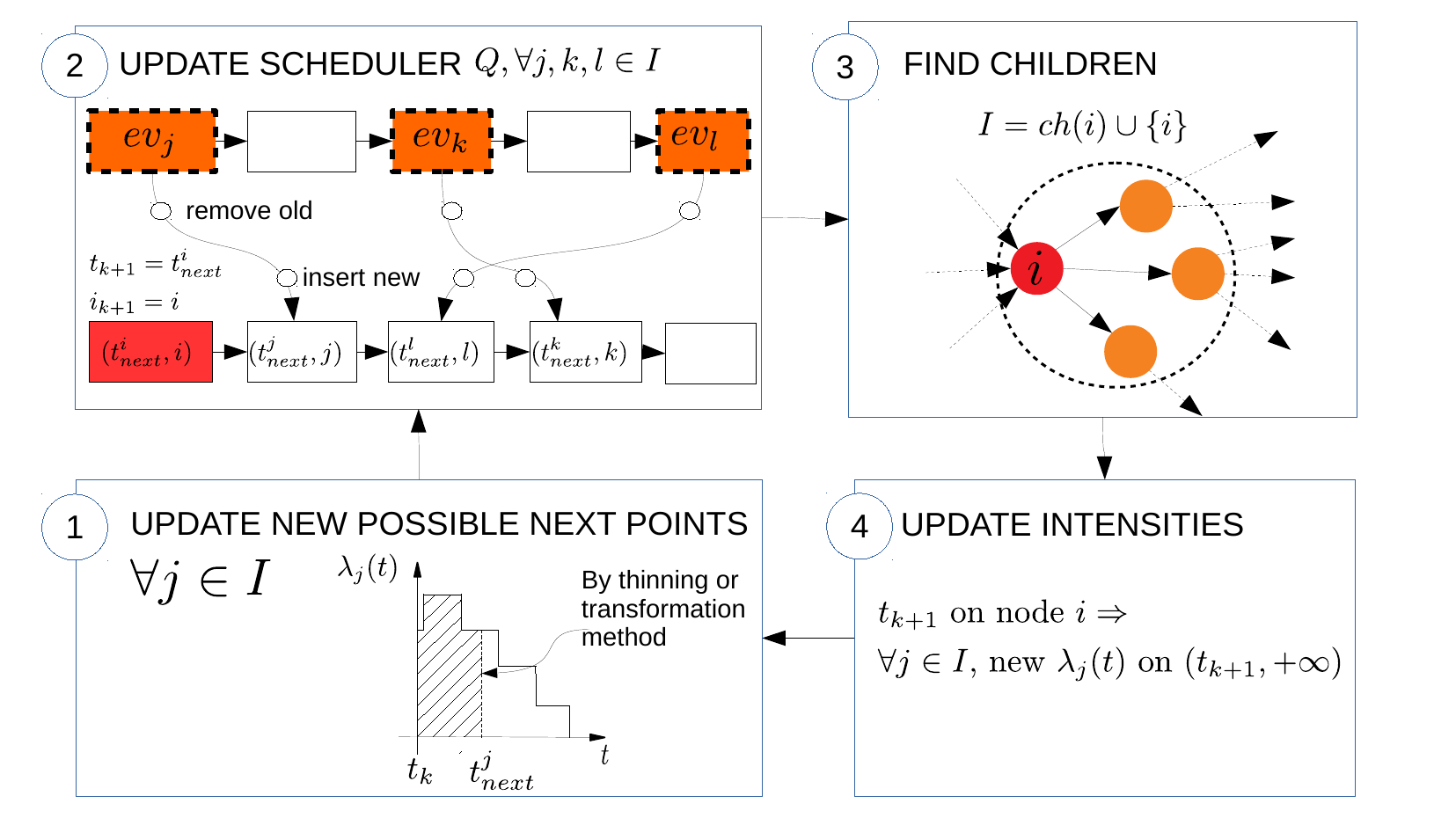}
            	\end{center}
                \caption{Steps of the local graph algorithm for the simulation of point processes. As for figure~\ref{fig:centra}, the intensities are piecewise constant.\label{fig:distrib}}
            \end{figure}

        	The local graph algorithm for point processes is described in Algorithm \ref{alg:distributed}. A visual representation is presented in Figure \ref{fig:distrib}. 
        	
        	Let us detail a bit more each step.
        	\begin{itemize}
        	    \item At Step a/, we  decide for the next possible point of each of the subprocesses $N_i$ by either using the transformation method or the thinning algorithm (Algorithm 1).
        	    \item At Step b/, for each $i\in I$ we remove the old event associated to $i$ and insert the one computed at Step a/.
        	    \item At Step c/, we retrieve the minimum of all possible times to find out which subprocess $i$ is actually generating a new point.
        	    \item  At Step d/, we look for the children of $i$ and update $I$.
        	    \item Step e/ depends hugely on the type of process at hand. The apparition of a new point has clear impact on the intensity but this depends on the formula. For instance if the intensity is given by Equation \eqref{eq:hawkes}, we need to shift intensities and update sums (see Section 5 for more details).
        	\end{itemize}
        	
        	More details about the algorithm steps are provided in Section~\ref{sec:Hawkes}, which presents in full details the application of the algorithm to the Hawkes case.
    
            \begin{algorithm}
                \begin{algorithmic}[1]
                    \State $k\leftarrow0, t_k \leftarrow 0$
                    \State $I \leftarrow \{ 1, ..., M \}$
                    \While{$t_k < T$}
                        \State \textbf{Compute the next possible points} $t^i_{next}$ for each $i \in I$ based on intensity $\lambda_{i}$ on $(t_k,+\infty)$
                        \State \textbf{Update} $Q$ with each next possible point $t^i_{next}$ for each $i \in I$\label{algline:dist:update-sched}
                        \State \textbf{Get next selected point } $t_{k+1} \leftarrow min\{t^i_{next} \}$ and $i$  the \textbf{associated node}, updating $Q \leftarrow Q^*$
                        \State \textbf{Find the children of $i$ and update} $\bm{I} \leftarrow ch(i) \cup \{ i \}$
                        \State \textbf{Update intensities } $\lambda_j(t)$ for each node $j \in I$ on  $(t_{k+1},+\infty)$
                  	\State $k\leftarrow k+1$ 
                    \EndWhile 
                    \State \textbf{return} $(t_1, ..., t_{k-1})$  points and associated nodes $(i_1, ..., i_{k-1})$
                \end{algorithmic}
                \caption{Local graph algorithm for the simulation of point processes: Application of the simulation activity tracking algorithm~\cite{cise-muzy}.}\label{alg:distributed}
                \AddNoteOneLine[black]{4}{4}{Step a/}
                \AddNoteOneLine[orange]{5}{5}{Step b/}
                \AddNoteOneLine[green]{6}{6}{Step c/}
                \AddNoteOneLine[purple]{7}{7}{Step d/}
                \AddNoteOneLine[blue]{8}{8}{Step e/}
            \end{algorithm}
        
	\section{Hawkes evaluation}\label{sec:Hawkes}
    	We want to evaluate the complexity of the previous algorithms, but this of course depends on the computational complexity of the conditional intensities associated to each point process. Previous general algorithms for simulating point processes are applied here to non explosive Hawkes processes with piecewise constant interactions with finite support (see Equation \eqref{eq:hawkes}). In this situation, note that the $\lambda_i$'s become piecewise constant, so that the complexity for calculating such intensities or updating them will be linked to the number of breakpoints of the corresponding piecewise constant function. Moreover with piecewise constant intensities, one can apply the transformation method directly, so we do not evaluate the complexity of the thinning /rejection step. The general algorithms are specified at data structure level in order to detail the computational complexity of each algorithmic step.

    	        \subsection{Notation and data structures}
            
    				
    				
    				
    				
    				
    				
    				
			
		    In the following, $\leftrightarrow$ means "corresponds to the mathematical notation". Data structure notation consists of:
	            
    			\begin{itemize}
    			    \item{$Q$:} A \emph{scheduler of next point events} $ev_i=(t_{next}^i,i)$, where $t_{next}^i$ is a \emph{possible next point} associated to node $i$.
    				\item{ $L[i] (\leftrightarrow \lambda_i)$: } is the scheduler  corresponding to the piecewise constant intensity of node $i$, The \emph{length of the scheduler} is $L_t^i = length(L[i])$ when $L[i][0].time=t$.  
    					  
    				
    				\item{$h[j][i] (\leftrightarrow h_{j \rightarrow i})$:} is the   scheduler corresponding to the piecewise constant interaction $h_{j \rightarrow i}$ (with support included in $[0,S]$) from node $j$ to node $i$. The \emph{maximum number of events in $h[j][i]$} is noted $A \geq length(h[j][i])$.
    				
    				
    				\item{$\overline{L}=L[1] \cup ... \cup L[M]$ ($ \leftrightarrow \overline{\lambda}(t)=\sum_{i=1}^M \lambda_i(t)$):} is a scheduler storing the piecewise sum of all $L[i]$, from node $1$ to node $M$ (cf. Figure \ref{fig:union}). 
    				
    				\item{$\overline{L}[i]=L[1]\cup...\cup L[i]$ ($ \leftrightarrow \overline{\lambda}_i(t)=\sum_{j=1}^i \lambda_j(t)$):} is a scheduler storing the partial piecewise  sum of the intensities from node $1$ to node $i$. Obviously, $\overline{L}[M]=\overline{L}$.
    			
    			\end{itemize}
    			
    	\subsection{Algorithm for the transformation method for piecewise constant intensities}
    	
    	Algorithm \ref{alg:geTnext}	presents the basic transformation method for piecewise constant conditional intensity functions, here represented by the scheduler $Q$ (cf. Equation \ref{eqrefinv}).
    	
    		  \begin{algorithm}
        		\begin{algorithmic}[1]
        		\Function{getTnext}{$Q$} 
        		    \State $V \sim \mathcal{U}[0,1]$
                    \State $integral \leftarrow 0$
                    \State $t_{next} \leftarrow Q[0].time$
                    \State $k \leftarrow 0$
                    \State $val\leftarrow 0$
                    \Repeat
                    \State $val\leftarrow val + Q[k].value$
                   
                    \State $integral \leftarrow integral + (Q[k+1].time - Q[k].time)\times val$
                     \State $k \leftarrow k+1$
                    \Until $(integral > -log(V) \text{ or } k=size(Q)-1)$
                    \If{$integral \leq -log(V) $}
                      \State $val \leftarrow val + Q[k].value$
                       
                    \EndIf 
                    \State \Return $t_{next} \leftarrow Q[k].time-\frac{integral+log(V)}{val}$ // $t_{next} \leftarrow +\infty$ if $val= 0$
                \EndFunction
        		\end{algorithmic}
        		\caption{Function $\textsc{getTnext(Q)}$ with $Q$ a scheduler storing the events corresponding to a piecewise constant intensity trajectory.}\label{alg:geTnext}
        	\end{algorithm}

        The complexity of the $\textsc{getTnext(Q)}$ operation is $\mathcal{O}(|Q|)$. 
        \subsection{Full scan and local graph algorithms}	
        
        Algorithm \ref{alg:centH} is the application of Algorithm \ref{alg:cent} to Hawkes processes. The mention to a, b c, d refers to the steps in Algorithm~\ref{alg:cent}. We split step d to lower the complexity. The value $T$ is an arbitrary stopping time, the condition $t_k<T$ ending the main loop in both Algorithm~\ref{alg:centH} and Algorithm~\ref{alg:distributedH} can be changed by whatever condition is deemed more appropriate by the modeller.
        
        	\begin{algorithm}[H]
        	\hspace*{\SpaceReservedForComments}{}%
            \begin{minipage}{\dimexpr\linewidth-\SpaceReservedForComments\relax}
        		\begin{algorithmic}[1]
        			\State $t_0 \leftarrow 0$
        			\State $k\leftarrow0$
        			\While{$t_k < T$}
            			\State
            			$\overline{L}[1] \leftarrow L[1]$ 
                			\ForAll{$j\in \{2,...,M\}$}
                			\State Prune intensities $\overline{L}[j] \leftarrow \overline{L}[j-1] \cup L[j]$
            			\EndFor
            			
            			\State $t_{k+1} \leftarrow \textsc{getTnext}(\overline{L}[M])$ 
            			
            			\ForAll{$j\in \{1,...,M\}$}
                			\State $L[j] \leftarrow L[j]^{t_{k+1}}_{pcw}$
             
        	            \EndFor
        	            
                	
                	\State $\ell[0]\leftarrow 0$
                	\ForAll{$j\in \{1,...,M\}$}
                	\State compute $\ell[j]=\overline{\lambda}_j(t_{k+1})$ \bf{ by } $\ell[j] \leftarrow \ell[j-1]+L[j][0].value$
                    \EndFor
                    \State Select the associated node $i_{k+1}$ as the only $j$ such that $\frac{\ell[j-1]}{\ell[M]}<V\leq\frac{\ell[j]}{\ell[M]}$ for  $V\sim \mathcal{U}[0,1]$
        	            \ForAll{$j\in \{1,...,M\}$}
        	               \State \textbf{Update intensities } $L[j] \leftarrow L[j] \cup h[i_{k+1}][j]_{\rightarrow_{t_{k+1}}}$
        	            \EndFor
 			          \EndWhile
            		\State $k \leftarrow k+1$
                	
        		    \State \textbf{return} points $(t_1, ..., t_{k-1})$ and associated nodes $(i_1, ..., i_{k-1}) $
        		\end{algorithmic}
        		\caption{Full scan algorithm for Hawkes processes}\label{alg:centH}
        		\end{minipage}
        		 \AddNote[black]{4}{6}{Step a/}
                \AddNoteOneLine[orange]{7}{7}{Step b/}
                \AddNote[green]{8}{9}{Step d1/}
                \AddNote[purple]{10}{13}{Step c/}
                \AddNote[green]{14}{15}{Step d2/}
        	\end{algorithm}
       
        \noindent Algorithm \ref{alg:distributedH} is the application of Algorithm \ref{alg:distributed} to Hawkes processes.
        
        	\begin{algorithm}[H]
        	 	\hspace*{\SpaceReservedForComments}{}%
              \begin{minipage}{\dimexpr\linewidth-\SpaceReservedForComments\relax}
                \begin{algorithmic}[1]
                    \State $I \leftarrow \{ 1, ..., M \}, k\leftarrow0$
                    \While{$t_k < T$}
                        \State $\textbf{Compute the next point } t^i_{next} \leftarrow \textsc{getTnext}(L[i])$ of each $i \in I$
                        \State $Q \leftarrow Q^* \oplus(t_{next}^i,i)$ of each $i \in I$
                        \State $t_{k+1} = Q[0].time$
                        \State $i_{k+1} = Q[0].value$
                        \State \textbf{Find children of $i$ and update} $I \leftarrow ch(i_{k+1}) \cup \{ i_{k+1} \}$
                        \ForAll{$i \in ch(i_{k+1})$}
                            \State $L[i] \leftarrow L[i]^{t_{k+1}}_{pcw} \cup h[i_{k+1}][i]_{\rightarrow_{t_{k+1}}}$
                  	    \EndFor
                  	    \If{$i \not\in ch(i_{k+1})$}
                  	        \State $L[i] \leftarrow L[i]^{t_{k+1}}_{pcw} \cup h[i_{k+1}][i]_{\rightarrow_{t_{k+1}}}$
                  	    \EndIf
                  	\State $k \leftarrow k+1$

                  	\EndWhile 
                  	\State \textbf{return} $(t_1, ..., t_{k-1})$  points and associated nodes $(i_1, ..., i_{k-1})$
                \end{algorithmic}
                \caption{Local graph algorithm for Hawkes processes} \label{alg:distributedH}
                \end{minipage}
        		 \AddNoteOneLine[black]{3}{3}{Step a/}
                \AddNoteOneLine[orange]{4}{4}{Step b/}
                \AddNote[green]{5}{6}{Step c/}
                \AddNoteOneLine[purple]{7}{7}{Step d/}
                \AddNote[blue]{8}{12}{Step e/}
            \end{algorithm}

        
        \subsection{Complexities of both algorithms}\label{sec:math}

        If $A$ (the number of breakpoints to describe the interaction functions $h_{j\to i}$) and $S$ (the support of the $h_{j\to i}$'s) are true constants, assumed to be of order $1$ in the sequel, the size of the different schedulers that are used in the previous algorithms are most of the time random and changing step after step. They depend in particular on the number of points of $N_j$ appearing in the interaction range that is $N_j([t-S, t))$. To evaluate further the order of such a random quantity, we know that a stationary Hawkes process has a mean intensity vector $m=(m_1,...,m_M)^T$ (see \cite{daley}):
        \begin{equation}\label{stat}
                m=(I_M-H)^{-1}\nu,
        \end{equation}
        with $\nu=(\nu_1,...,\nu_M)^T$, $I_M$ the identity matrix of size $M$ and $H=(\int_0^{+\infty} h_{j\to i}(x) \dd x)_{i,j=1,...,M}$. Note that the linear Hawkes process is explosive when the spectral radius of $H$ is strictly larger than $1$ (we refer the reader to~\cite{bremaudStability} and~\cite{mastromatteo} for demonstrations even in the non-linear cases). When the spectral radius is strictly less than $1$, the non explosive  Hawkes process, with no points before time $0$, has always less points than the stationary version. Therefore,  
        \begin{equation}\label{upperNj}
                \mathbb{E}(N_j([t-S, t))) \leq m_j S
        \end{equation}
        with $m_j$ given by Equation \eqref{stat}.
            
        Moreover, the local independence graph for Hawkes process is completely equivalent to the graph with edge $j\to i$ if and only if $h_{j\to i}$ is non zero. The corresponding adjacency matrix is denoted $R=(\textbf{1}_{\int h_{j\to i}\neq 0})$.

        At time $t$, the scheduler $L[i]$ describes the piecewise constant conditional intensity  $\lambda_i(.)$ on $[t,+\infty)$ in absence of new points after $t$. The number of breakpoints of $L[i]$ is denoted $L^i_t$.
        But~\eqref{eq:hawkes} can be rewritten as
        \[
                \lambda_i(t)=\nu_i + \sum_{j\in pa(i)} \sum_{T\in N_j, T \in [t-S, t)}  h_{j\to i}(t-T)
        \]
       So we can first note that $L^i_t$ and therefore its expectation $\mathcal{L}_i= \mathbb{E}(L^i_t)$ are always larger than $1$ because the scheduler $L[i]$ is at least of size $1$. Moreover this  piecewise constant function has potential breakpoints at  all $T+a$, for $T\in N_j, T \in [t-S, t)$, and $a$ breakpoints of $h_{j\to i}.$

        \noindent Therefore we can compute the order of magnitude of $L^i_t$, which is the length of the scheduler associated to $\lambda_i(t)$, by
        \[
                L^i_t =\mathcal{O}\left(1+A \sum_{j\in pa(i)}  N_j([t-S, t)\right)
        \]
        where $\mathcal{O}$ means that there exists an absolute positive constant $C$ such that $$
        L^i_t \leq C \left(1+A\sum_{j\in pa(i)} N_j([t-S, t) \right).
        $$
      
        \noindent In expectation, this gives, thanks to \eqref{upperNj} and since $AS = \mathcal{O}(1)$, 
    \begin{align}\label{calL}
                  \mathcal{L} &=\mathcal{O}\left(1+Rm\right)=\mathcal{O}\left( 1+R(I_M-H)^{-1}\nu\right)
    \end{align}
        with $\mathcal{L}=(\mathcal{L}_i)_{i=1,...,M}$, the notation $\mathcal{O}$ being understood coordinate by coordinate.
        
        Now we can evaluate the (mean) complexity of both algorithms, replacing $L^i_t$ by $\mathcal{L}_i$ thanks to the respective complexities of each operation on the schedulers, see Section  \ref{sec:sched}.
        
    \paragraph{Full-Scan Algorithm}
    Step a/ has a complexity of 
    $$\mathcal{O}\left(\sum_{j=1}^M \mathcal{L}_i \log(\sum_{i=1}^j \mathcal{L}_i)\right)=\mathcal{O}\left( |\mathcal{L}|_1 \log|\mathcal{L}|_1\right)$$
    with $|\mathcal{L}|_1=\mathcal{L}_1+...+\mathcal{L}_M \geq M$
     Step b/ has complexity $\mathcal{O}\left( |\mathcal{L}|_1\right)$, as well as Step d1/.
    Step c/ has complexity $\mathcal{O}(M)\leq \mathcal{O}\left( |\mathcal{L}|_1\right)$.
Step d2/ has complexity
   $$\mathcal{O}\left(\sum_{j=1}^M (A \log(\mathcal{L}_i)+A+\mathcal{L}_i)\right)=\mathcal{O}\left( |\mathcal{L}|_1 \log|\mathcal{L}|_1\right)$$
   
   So globally one iteration of the full scan algorithm has a complexity of the order
   $$
   \mathcal{O}\left( |\mathcal{L}|_1 \log|\mathcal{L}|_1\right)=\mathcal{O}\left( (M+|Rm|_1) \log(M+|Rm|_1)\right).
   $$
    Therefore since the mean total number of iterations of this algorithm is also the mean total number of points produced on $[0,T]$, that is $T |m|_1$, the full-scan algorithm should have  the following mean complexity
    \begin{equation}\label{full}
    \mathcal{O}\left( T|m|_1(M+|Rm|_1) \log(M+|Rm|_1)\right).
    \end{equation}
    As expected, the complexity is linear with the duration $T$ of the simulation.
    Moreover  this complexity heavily depends on the whole set of parameters (type of graph, strength of the interaction functions etc), because in particular these parameters affect the number of points that have to be produced. So for very unbalanced networks where $|m|_1=\mathcal{O}(1)$ (if for instance only one node in the whole network is clearly active and the others almost silent), the complexity seems to be of order $\mathcal{O}(TM\log(M))$. But these very unbalanced networks are not the most usual.
    Let us look now at more balanced networks. Let us assume that all the $m_j$'s are roughly the same and are of order $1$ (no really small $m_j$) and that the number of parents of a given node is bounded by $d$, this give us a complexity of 
    $$ \mathcal{O}\left(TM^2 d \log(dM)\right).$$
    So up to the log factor, if the network is sparse but balanced, the complexity is quadratic in the number of nodes of the network. If the network is a full complete graph, the complexity is cubic in $M$.
    
    \paragraph{Local graph algorithm}
    As before we need to evaluate first the complexity of one iteration of the algorithm. But because $I$ is chosen at step d/ and the size of $I$ impacts the complexity of steps a/b/ and e/, we choose to evaluate the complexity of an iteration which starts with e/ and then does a/b/ c/ and d/, so that that until d/ the set $I$ is the same.
    
    \noindent If the node $i_{k+1}=j$, then the complexity of step e/ is
    $$\mathcal{O}\left( \mathcal{L}_j+\sum_{i\in ch(j)} ( \mathcal{L}_i+A+A\log(\mathcal{L}_i))\right)=\mathcal{O}\left( \mathcal{L}_j+\sum_{i\in ch(j)} ( \mathcal{L}_i+\log(\mathcal{L}_i))\right).$$
    The complexity of step a/ is $$\mathcal{O}\left( \mathcal{L}_j+\sum_{i\in ch(j)} \mathcal{L}_i\right).$$
    The complexity of step b/ is
    $$\mathcal{O}\left( \log(M)+\sum_{i\in ch(j)} \log(M)\right).$$
    Steps c/ and d/ have complexity $\mathcal{O}(1)$.
    
    So for one iteration "e/a/b/c/d/" after a point on node $j$, the complexity is
    $$\mathcal{O}\left( \mathcal{L}_j+\log(M)+[R'(\mathcal{L}+\log(M){\bf 1})]_j\right),$$
    with $R'$ the transpose of $R$ and ${\bf 1}$ the vector of size $M$ full of ones.
    
    The main point is that a point in $N_j$ is appearing in average at most only $Tm_j$ times during the simulation since $m_j$ is the mean intensity of  $N_j$, which leads us to a global complexity of
    \begin{equation}\label{local}
    \mathcal{O}\left(T m'\mathcal{L} +T \log(M) |m|_1+ T m'R'[\mathcal{L}+\log(M) {\bf 1}]\right)=T\mathcal{O}\left(m'Rm+\log(M)|m|_1+m'R'Rm+\log(M)|Rm|_1\right).
    \end{equation}
    As before this is linear in the duration of the simulation $T$ and depends heavily on the parameters. But the complexity is much lower. Indeed, for very unbalanced networks where only one node is really active, the complexity
    logarithmic in $M$. For balanced networks where the $m_j$'s are roughly the same and if the number of children of a given node, as well as the number of parents is bounded by $d$, then we get a complexity of
    $$\mathcal{O}\left(T M d[d+\log(M)]\right),$$
    For sparse balanced graphs, we therefore get a complexity which is linear in $M$ up to logarithmic factors. The gain is clear with respect to the full scan algorithm. For complete graphs, we also get a cubic complexity in terms of $M$, as the full scan algorithm but without logarithmic factors. 
    
    So at least theoretically speaking, it seems that the local graph algorithm is always a better choice than the full-scan algorithm, with a clear decrease of complexity from quadratic to linear in the number of nodes for balanced sparse graphs.
    
\section{Numerical experiments}\label{sec:sim}    
This section is devoted to two main problems: statistically proving that both algorithm (full scan and local graph) indeed simulate a Hawkes process and asserting that  the local graph algorithm clearly outperforms the full scan algorithm.

\subsection{Hardware and software specifications}
The main simulations have been performed on 5 nodes of a Symmetric MultiProcessing (SMP), i.e., shared memory, computer1. Each of this computational nodes has up to 20 physical cores (2\*10), 25 MB of cache memory and 62.5 GB of RAM. The processors are Intel(R) Xeon(R)CPU E5-2670 (v0 and v2) at 2.60 GHz. The statistical analysis required more RAM, so we used another type of node, which has 770GB of RAM, 25MB of cache memory, 20 physical cores (2\*10), each processor being an Intel(R) Xeon(R)CPU E5-2687W v3 at 3.10GB. The algorithms were implemented in C++ programming language (2011 version). No other external libraries were used for the simulator, which is compiled using gcc 4.7. The plots and statistical analyses were obtained using using R software (v3.6), part of it using the UnitEvent package (v0.0.5).


\subsection{Statistical analysis}
 We generated an Erd\"os-R\'enyi network of 100 nodes with connection probability $p=1/100$, that is fixed for the rest of the statistical analysis. When an edge $j\to i$ is in the graph, we associate it to an interaction function $t \mapsto h_{j\to i}(t)=5\cdot\mathbx{1}_{t\in[0,0.02]}$. The spontaneous parameters $\nu_i$ are all fixed to 10. Out of this multivariate Hawkes process, we focus on two nodes $a$ and $b$. The node $a$ is fully disconnected, meaning the corresponding process should be an homogeneous Poisson process of rate 10. The node $b$  is the one with the largest number of parents (4 parents).
 
 \paragraph{Time transformation}
 In \cite{Ogata_test}, Ogata derives methodological benchmarks to assess if the data are obeying a  point process with a given intensity, and in particular Hawkes processes. This is based on the time-rescaling theorem (see for instance \cite{time-rescale}), which says that if $\lambda_s$ is the conditional intensity of the point process $N$ and if $\Lambda(t)=\int_0^t \lambda_s ds$, then the points $\tilde{N}=\{\Lambda(T), T\in N\}$ form an homogeneous Poisson process of rate 1. Ogata proposed the following method to test that a given point process has intensity given by $\lambda_s$
 \begin{itemize}
     \item Apply the time-rescaling transformation. This leads to a point process $\tilde{N}$.
     \item Test that the consecutive delays  between points of $\tilde{N}$ obeys an exponential distribution of rate 1, for instance by Kolmogorov-Smirnov test ({\bf Test 1})
     \item Test that the points of $\tilde{N}$  themselves are uniformly distributed,  for instance by Kolmogorov-Smirnov test ({\bf Test 2}).
     \item Test that the delays between points of $\tilde{N}$ are independent, for instance by checking that the autocorrelation between delays with a certain lag are null ({\bf Tests 3}). We performed them up to lag 9.
\end{itemize}
We simulated the multivariate Hawkes process on $[0,T]$ with $T=150$ and we applied the previous tests to node $a$ and node $b$.

\begin{figure}
    \centering
    \includegraphics[width=\textwidth]{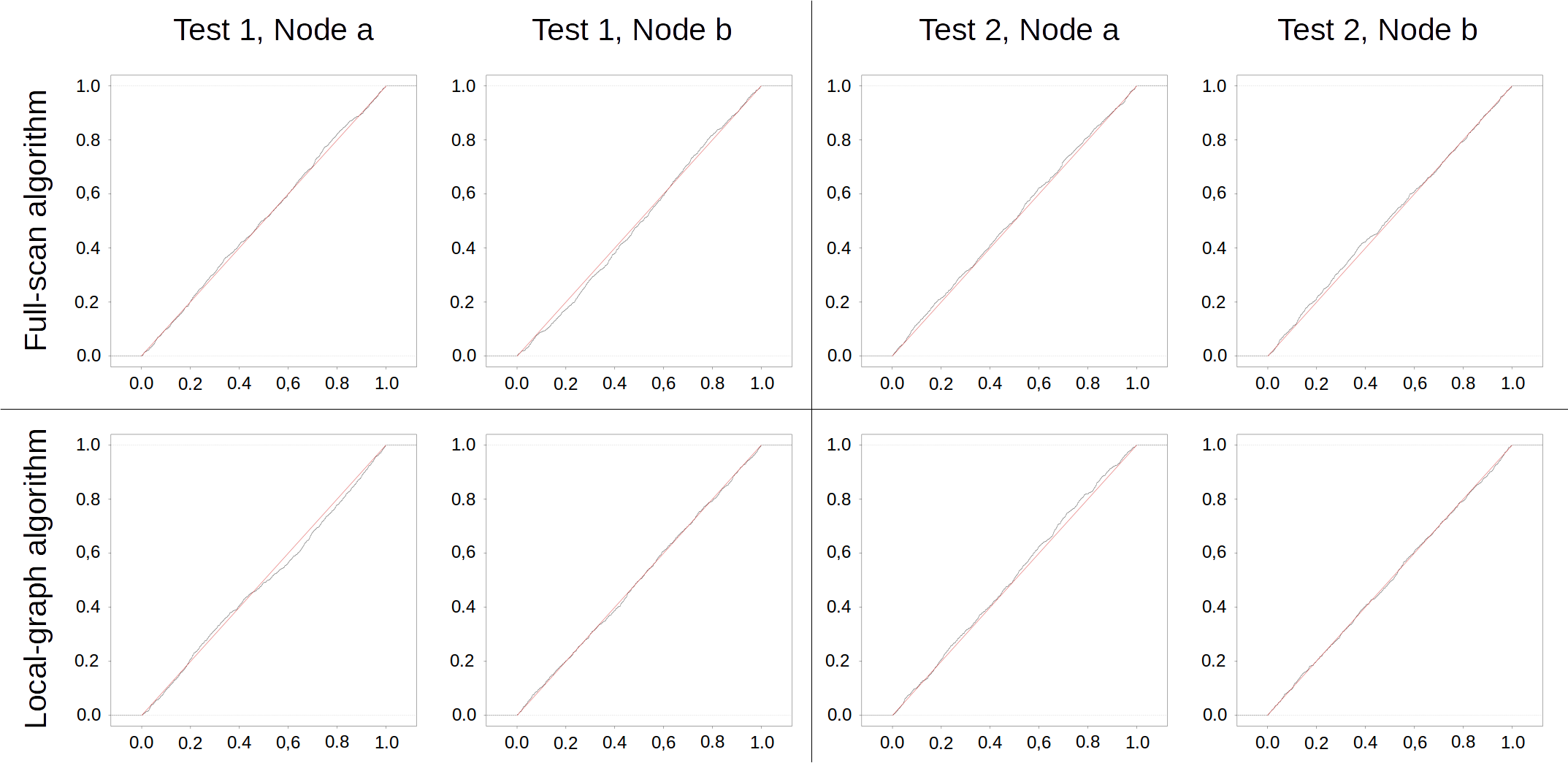}
    \caption{Cumulative distribution functions of the p-values of Test 1 and 2. In columns the test and node, in rows the algorithms (full scan then local graph)}\label{fig:p-values}
\end{figure}

\begin{figure}
    \centering
    \includegraphics[width=\textwidth]{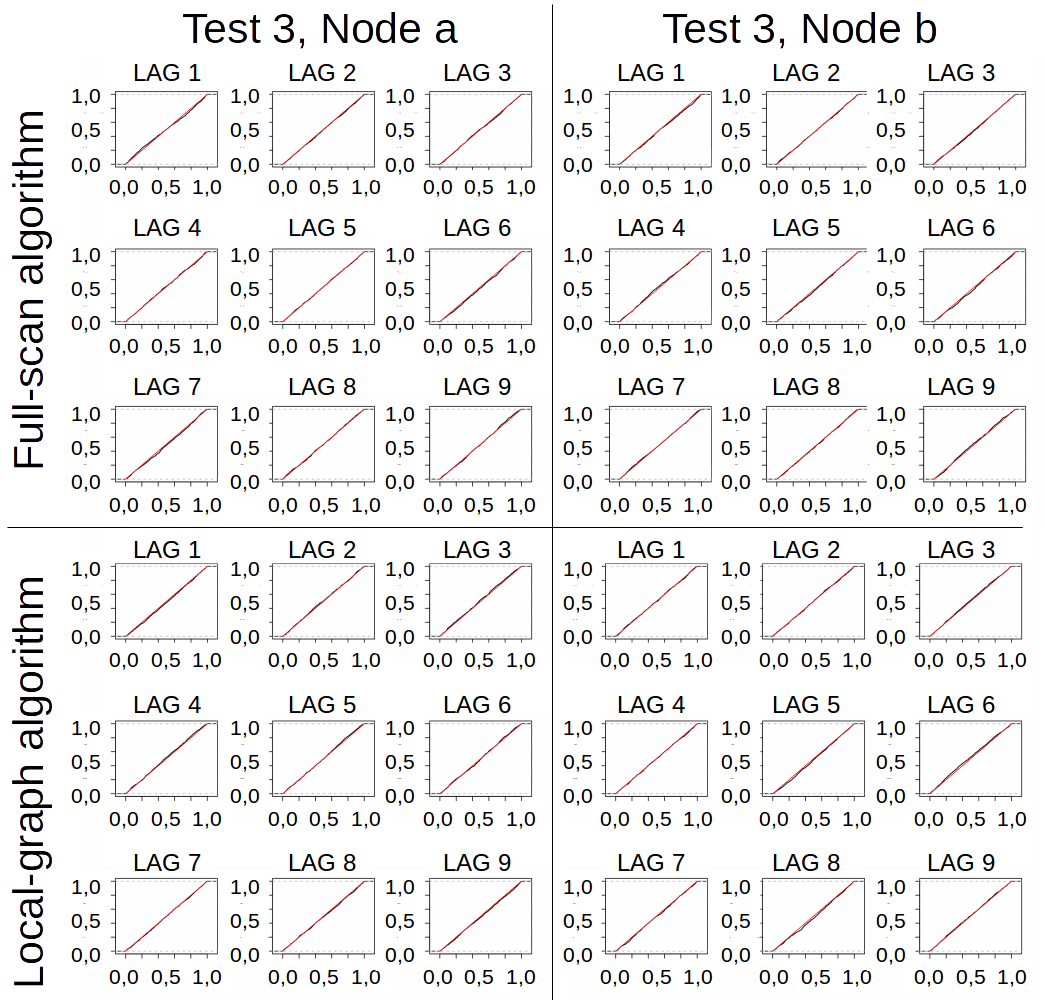}
    \caption{Cumulative distribution functions of the p-values of Test 3. In columns the node, in rows the algorithm (full scan then local graph)}\label{fig:acf-p-values}
\end{figure}

 \begin{table}[H]
    \caption{Table of the p-values of a Kolmogorov-Smirnov test (for uniformity) applied to the p-values obtained with tests 1, 2 and 3 for $1000$ independent simulations of the same Hawkes point processes (with the same underlying graph).}\label{tab:ksofks}
    \centering
    \begin{tabular}{c|c|c|c|c}
        & \multicolumn{2}{c}{\textsc{full-scan}} & \multicolumn{2}{c}{\textsc{local-graph}}\\\hline
        & Node $a$ & Node $b$ & Node $a$ & Node $b$  \\\hline
        \textbf{Test 1}&0.5384525&0.1491268&0.0594925&0.86789804\\
\textbf{Test 2}&0.6008973&0.2462138&0.1819709&0.99025263\\
\textbf{Test 3} with lag 1 &0.1602718&0.1781804&0.4385096&0.92162419\\
\textbf{Test 3} with lag 2 &0.7498109&0.9038829&0.6954876&0.90558993\\
\textbf{Test 3} with lag 3 &0.5604420&0.7220130&0.4144515&0.77140051\\
\textbf{Test 3} with lag 4 &0.7003987&0.1838913&0.4367523&0.83833821\\
\textbf{Test 3} with lag 5 &0.9960351&0.4009543&0.3740874&0.14749913\\
\textbf{Test 3} with lag 6 &0.1883506&0.1246654&0.4387684&0.12202262\\
\textbf{Test 3} with lag 7 &0.1259022&0.8588754&0.9114556&0.47030751\\
\textbf{Test 3} with lag 8 &0.8848928&0.9720601&0.5200698&0.03765871\\
\textbf{Test 3} with lag 9 &0.2278844&0.3880436&0.5042846&0.92768290\\
    \end{tabular}
\end{table}
If we have simulated indeed the correct Hawkes processes for the processes associated to node $a$ and $b$, the p-values should be uniform. So we performed $1000$ simulations of the same Hawkes process (with the same underlying graph) but with different pseudorandom generator seeds for the simulation of the points themselves. We can visually check that they are indeed uniform by seeing diagonals for their cumulative distribution functions (see Figures~\ref{fig:p-values} and \ref{fig:acf-p-values}). In order to confirm this qualitative result with a more quantitative one, the p-values for the three tests 1, 2 and 3 are independently tested for uniformity with another Kolmogorov-Smirnov test. The resulting p-values are displayed in Table~\ref{tab:ksofks}. 

\paragraph{Martingale properties}

\begin{figure}
     \centering
     \includegraphics[width=\textwidth]{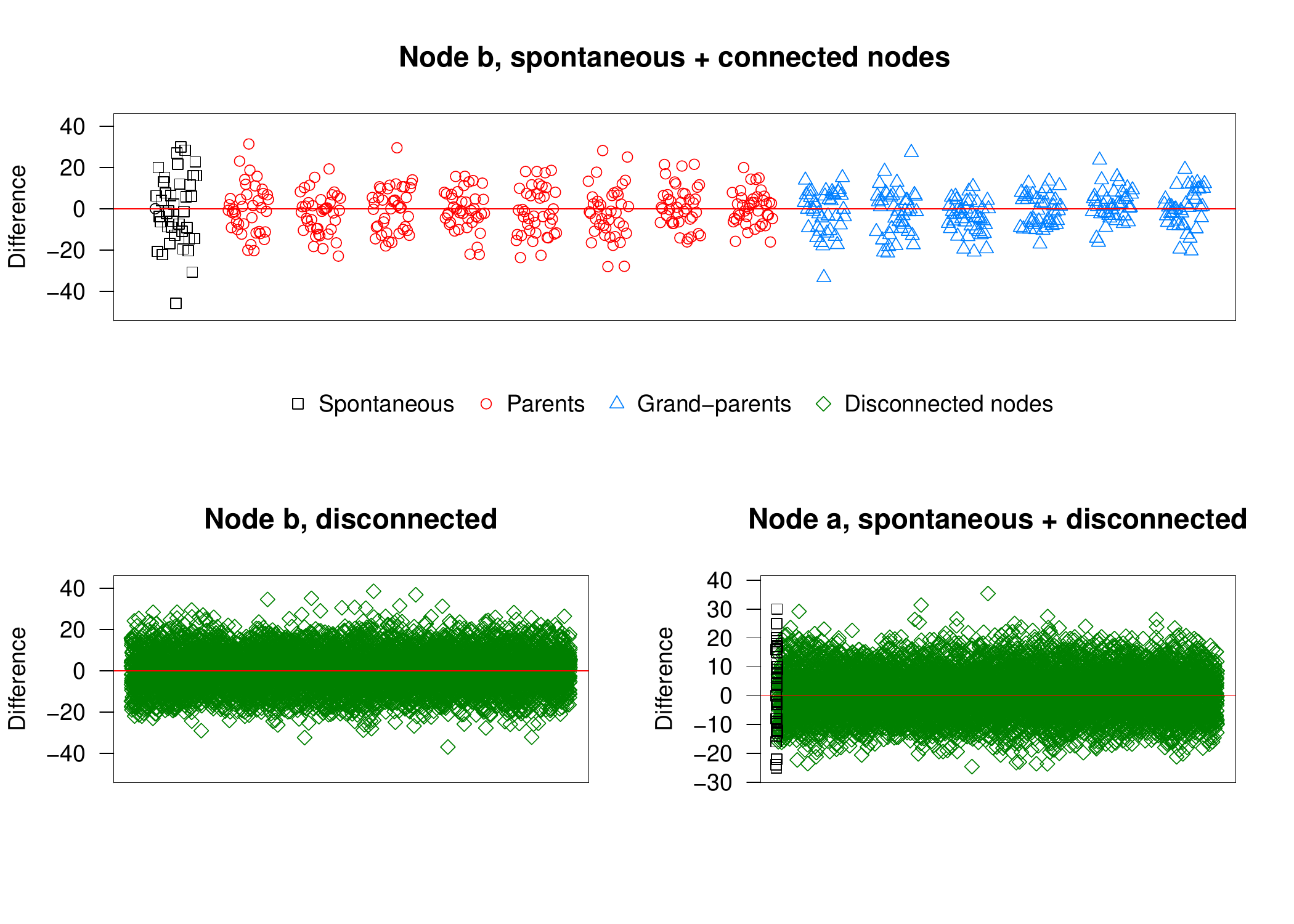}
     \caption{Full scan algorithm: verifying the Martingale property for Nodes a and b. The black points represent the $X^1$ (spontaneous), then for the nodes connected to Node b, $X^{2j}$ and $X^{2j+1}$ are displayed in red. In blue are the $X^k$ and $X^{k+1}$ (still for Node b) from Node b's grand-parents to Node b's parents. Finally the two green scatter plots show the Nodes not disconnected from b and a respectively.}
     \label{fig:centrees-full-scan}
 \end{figure}

\begin{figure}
     \centering
     \includegraphics[width=\textwidth]{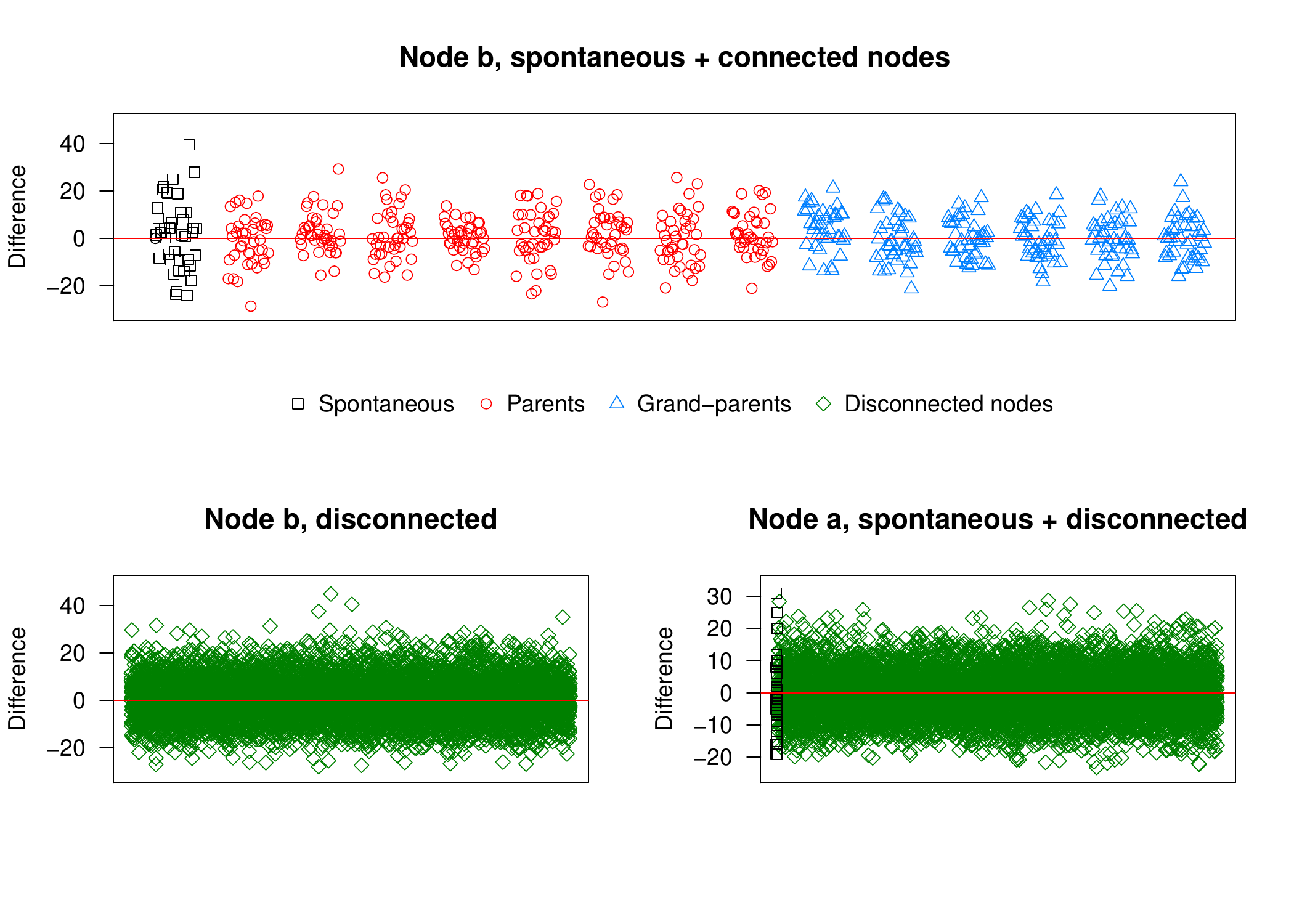}
     \caption{Local-graph algorithm: verifying the Martingale property for Nodes a and b. The black points represent the $X^1$ (spontaneous), then for the nodes connected to Node b, $X^{2j}$ and $X^{2j+1}$ are displayed in red. In blue are the $X^k$ and $X^{k+1}$ (still for Node b) from Node b's grand-parents to Node b's parents. Finally the two green scatter plots show the Nodes not disconnected from b and a respectively.}
     \label{fig:centrees-local-graph}
 \end{figure}

 Another very important property of the Hawkes process is that $t\mapsto N_t-\Lambda_t$ is a martingale and this property remains true if we integrate with respect to a predictable process. So for each node $a$ or $b$, we can compute
 $$X^k=\int_0^T \psi^k_t (dN_t-d\Lambda_t),$$
 for $\psi^1_t=1$ or $\psi^{2j}_t=N_j([t-0.02,t)]$ or $\psi^{2j+1}_t=N_j([t-0.04,t-0.02)]$.
 If the martingales properties are true, then the variable $X^k$ for each $k$ should be centered around 0. We also expect eventually different behaviors, when $k=1$, which corresponds to the spontaneous part or when $k=2j$ or $2j+1$ for a node $j$ which is connected to the node of interest or disconnected from the node of interest.
 We simulated the network $40$ times on $[0,T]$ with $T=20$ and reported the $X^k$.
 We see on Figure~\ref{fig:centrees-full-scan} and \ref{fig:centrees-local-graph} that the variables $X^k$ are indeed centered in both cases as expected. So we can conclude  that  both algorithms indeed simulate the given Hawkes process.
 
 \subsection{Performance}
  We want to assess the performances of both algorithm in the main interesting case: sparse balanced networks. To do so, we took three different topologies of graphs:
  \begin{itemize}
	 \item Erd\"os-R\'enyi: a topology model where each edge has a  probability $p$ of being present or absent, independently of the other edges. We took $p$ in $\{0, \frac{1}{M}, \frac{2}{M}, \dots, \frac{(ln(M)-1)}{M}\}$ for $M$ the number of nodes. These choices for $p$ ensure a sparse graph with roughly speaking $d=pM$ parents and children for each node.
	 \item Cascade: a classical topology model where each node has exactly one parent and one child (except for two nodes, start and end, that have respectively no parent and one child, and one parent and zero child), and there are no cycles in the network. There is only one graph per number $M$.
	\item Stochastic-Block: it is an Erd\"os-R\'enyi by block. In our setting the nodes are partitioned in two blocks and the matrix gives the probability of inter-block connection of intra-block connection (see Table \ref{tab:stochastic-bloc-topologies}).
\end{itemize}
  \begin{table}[H]
                    \centering
                    \caption{The three bloc sizes vectors (first line) and the three probability matrices (second line) used for the simulations.}
                    \label{tab:stochastic-bloc-topologies}
                    \begin{tabular}{l|ccc}
                        Block sizes & $\begin{pmatrix}\frac{M}{2} & \frac{M}{2}\end{pmatrix}$&
                        $\begin{pmatrix}\frac{M}{2} & \frac{M}{2}\end{pmatrix}$&
                        $\begin{pmatrix}\ln{M} & M-\ln{M}\end{pmatrix}$\\
                        
                        Probability matrices & $\begin{pmatrix}\frac{2}{M}\ln\frac{M}{2} & 0\\
                            0 & \frac{2}{M}\ln\frac{M}{2}\end{pmatrix}$ &
                        $\begin{pmatrix}0 & \frac{2}{M}\ln\frac{M}{2}\\
                            \frac{2}{M}\ln\frac{M}{2} & 0\end{pmatrix}$ &
                        $\begin{pmatrix}0 & \frac{\ln(\ceil{\ln M})}{\ceil{\ln M}}\\
                        \frac{\ln(M-\ceil{\ln M})}{(M-\ceil{\ln M})} & 0\end{pmatrix}$
                    \end{tabular}
                \end{table}
  Each graph was generated using a different pseudorandom generator seed. Each existing edge $j\to i$ is associated with an interaction function $t \mapsto h_{j\to i}(t)=5\cdot\mathbx{1}_{t\in[0,0.02]}$. We computed the largest eigen-value of the corresponding matrix $H$. If it is larger than $1$, this graph should be discarded.
  To force the balance of the network, we decided to take $m=(10,...,10)$ and compute the $\nu_i$'s by $\nu= (I_M-H)m$. It may happen that some of the $\nu_i$'s become negative. These graphs should be also discarded, as by construction the conditional intensity $\lambda(t)$ of a Hawkes point process must remain positive at all time, and so it is the case for the background intensity $\nu_i$ (see \cite{bremaud1981point} for reference).
  Because of the parameter values, especially the interaction functions, no graph was discarded here. A total of $2890=1770+280+840$ (Erd\"{o}s-R\'enyi + Cascade + Stochastic-Block) graphs was obtained with $M=\{10, 20, \dots, 100\}\cup\{150, 200, \dots, 500\}\cup\{600, 1100, \dots, 5100\}$ for the local-graph algorithm, and $M=\{10, 20, \dots, 100\}\cup\{150, 200, \dots, 500\}$ for the full-scan algorithm. Once the parameters of the Hawkes process are fixed, we simulated 10 times each process on $[0,T]$ with $T=10$, each simulation with a different generator seed.

\begin{figure}[H]
        \includegraphics[width=\linewidth]{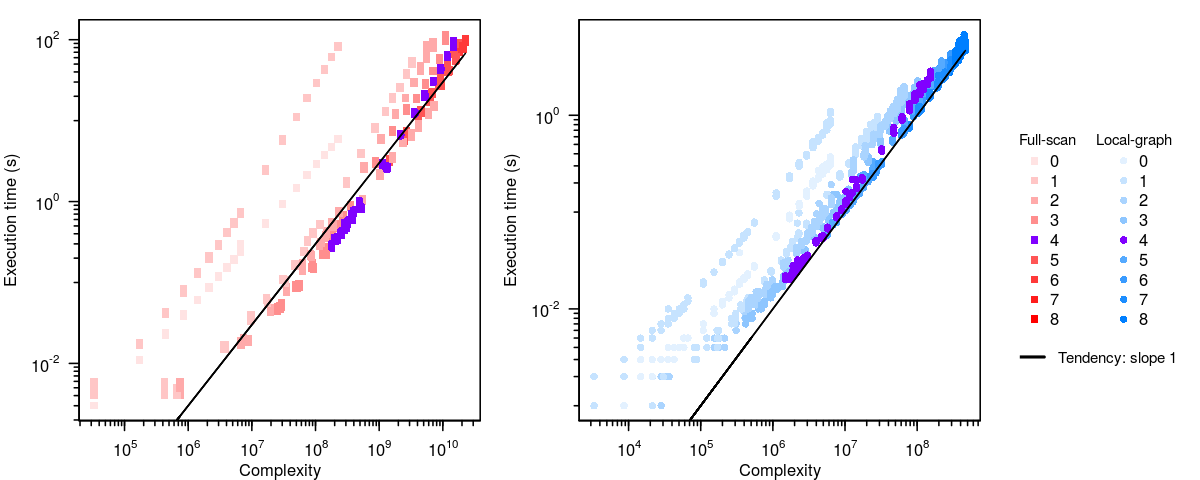}
        \caption{Three topologies together: the execution time (vertical axis, log-scaled) and the theoretical complexity (horizontal axis, log-scaled), for the full scan algorithm (red squares, left part) and the local-graph algorithm (blue circles, right part). A line of slope $1$ is displayed in black, on both scatter plots, showing the equivalence. The colour gradient represents similar values of the mean number of connections per process. A particular value (mean of $4$ children per process) is emphasised with a violet tone. The number of nodes is $M=\{10, 20, \dots, 100\}\cup\{150, 200, \dots, 500\}$ for the full-scan algorithm and $M=\{10, 20, \dots, 100\}\cup\{150, 200, \dots, 500\}\cup\{600, 1100, \dots, 5100\}$ for the local graph algorithm.}\label{fig:simtheo}
    \end{figure}

    
 Figure\ref{fig:simtheo} shows that the theoretical complexities of both full scan and local graph algorithms are equivalent to their actual execution times.

\begin{figure}[H]
      \centering
      \includegraphics[width=\textwidth]{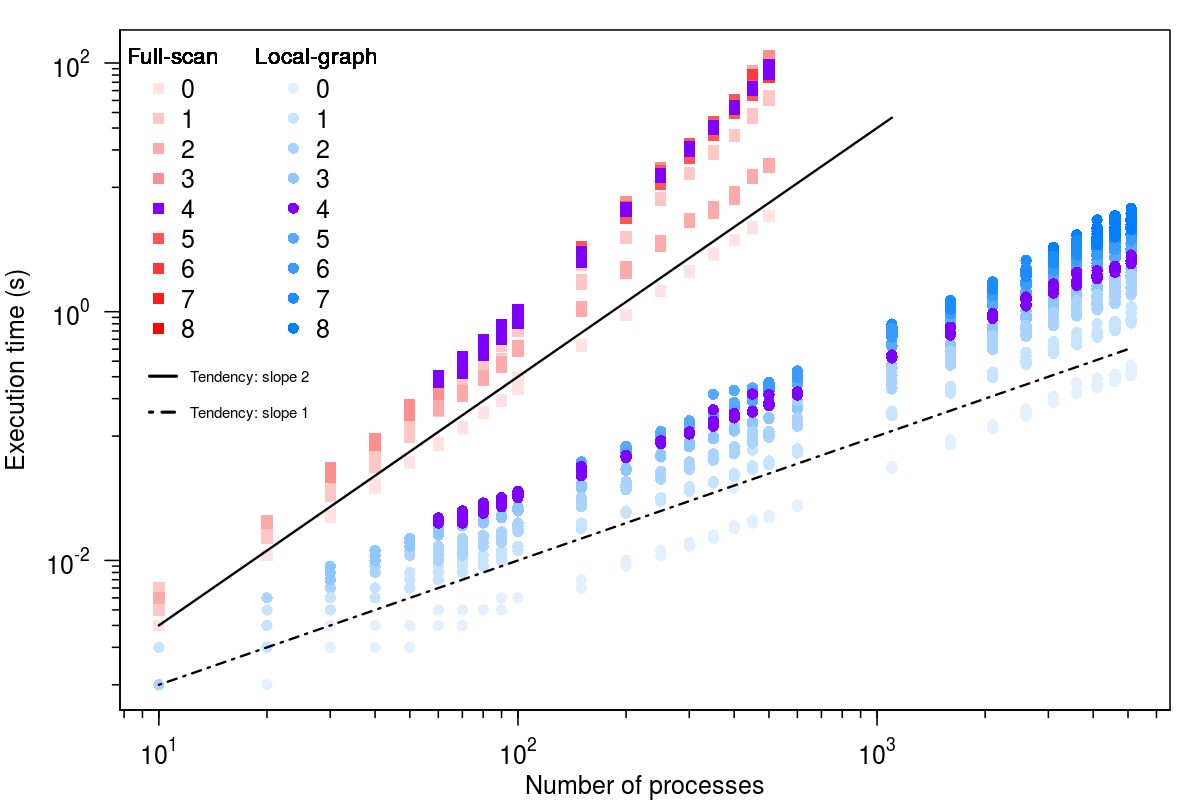}
      \caption{Three topologies together:  the execution time (vertical axis, log-scaled) and the theoretical complexity (horizontal axis, log-scaled), for the full scan algorithm (red squares, left part) and the local-graph algorithm (blue circles, right part). A  line of slope 2 is displayed in black, on the scatter plot for the full scan algorithm, and a line of slope 1 for the local graph algorithm. The colour gradient represents similar values of mean numbers of connections per process. A particular value (mean of 4 children per process) is emphasised with a violet tone. The number of nodes is $M=\{10, 20, \dots, 100\}\cup\{150, 200, \dots, 500\}$ for the full-scan algorithm and $M=\{10, 20, \dots, 100\}\cup\{150, 200, \dots, 500\}\cup\{600, 1100, \dots, 5100\}$ for the local graph algorithm.}\label{fig:mm2sim}
  \end{figure}

  Figure~\ref{fig:mm2sim} shows that the execution time is quadratic for the full scan algorithm and linear behaviour for the local graph algorithm. For example, when the local graph algorithm is executed in less than $10$s for more than $5000$ nodes, the execution of the full scan algorithm takes about $100$s for $500$ nodes. The local graph algorithm clearly outperforms the full scan algorithm.
  
    \section{Conclusion}
	We presented a new discrete event simulation for point processes: 
	the local graph algorithm, aiming at tracking only the nodes changing state in the network, only updating their children (based on the local independence graph hypothesis \cite{didelez}). The computational complexity reduction of the local graph algorithm with respect to the full scan algorithm (an adaptation of Ogata's algorithm \cite{ogata}) is from $M^2$ to $M$. Although there was no simulation algorithm able to simulate large point process networks, the local graph algorithm now opens new perspectives for simulating such networks. Especially, based on the local graph generation, an interesting perspective concerns the memory reduction. Instead of statically storing the whole network topology in memory at the beginning of the simulation, only the local graphs corresponding to the children of changing state nodes could be dynamically generated during the simulations \cite{refId0}. Based on time asynchrony, this corresponds to asynchronous dynamic structure changes in discrete event systems \cite{dyndevs}. Generating only local graphs with respect to the whole network graph should allow simulating very large networks. The same complexity reduction order is expected. However, at execution time level, the cost of re-generating the local graphs will have to be taken into account.
	
	The network structure plays a central role in the arguments. While we assume that all processes in the population are of the same type, the connectivity between the processes in the population is not homogeneous. Each process in the population of $N$ nodes receives input from $C$ randomly selected processes in the population. Sparse connectivity means that the ratio $\delta=\frac{C}{N}\ll 1$ is a small number. One can ask if it is this realistic. In the context of the human brain, a typical pyramidal neuron in the cortex receives several thousand synapses from presynaptic neurons while the total number of neurons in the cortex is much higher~\cite{PAKKENBERG200395}. Thus globally the cortical connectivity $\frac{C}{N}$ is low. On the other hand, we may concentrate on a single column in visual cortex and define, e.g., all excitatory neurons in that column as one population. We estimate that the number $N$ of neurons in one column is below ten thousand. Each neuron receives a large number of synapses from neurons within the same column. In order to have a connectivity ratio of 0.1, each neuron should have connections to about a thousand other neurons in the same column. \cite{gerstner2002spiking}. In the brain, last estimations consist of $86$ billions of neurons~\cite{nbNeurons}, each neuron having around $7'000$ connections. \emph{Either for the overall brain or for a single column of the visual cortex, the hypothesis of sparse connectivity of the network remains valid}. This work thus  allows achieving grounded stochastic simulations of the neuronal functional interactions in parts of the human brain.

\section*{Acknowledgement}
  For the SMP simulations we would like to deeply thank the LIMOS CNRS laboratory, from the University of Clermont-Auvergne, which graciously provided access. In particular we would like to thank their current administrators, Hélène Toussaint, William Guyot-Lénat and Boris Lonjon, for their valuable help.\\
  This work was supported by the French government, through the $\text{UCA}^\text{Jedi}$ and 3IA C\^ote d'Azur Investissements d'Avenir managed by the National Research Agency (ANR-15- IDEX-01 and ANR-19-P3IA-0002) and by the interdisciplinary Institute for Modeling in Neuroscience and Cognition (NeuroMod) of the Universit\'e C\^ote d'Azur.

	\bibliographystyle{ACM-Reference-Format}
	\bibliography{biblio}


\begin{thebibliography}{35}


\ifx \showCODEN    \undefined \def \showCODEN     #1{\unskip}     \fi
\ifx \showDOI      \undefined \def \showDOI       #1{#1}\fi
\ifx \showISBNx    \undefined \def \showISBNx     #1{\unskip}     \fi
\ifx \showISBNxiii \undefined \def \showISBNxiii  #1{\unskip}     \fi
\ifx \showISSN     \undefined \def \showISSN      #1{\unskip}     \fi
\ifx \showLCCN     \undefined \def \showLCCN      #1{\unskip}     \fi
\ifx \shownote     \undefined \def \shownote      #1{#1}          \fi
\ifx \showarticletitle \undefined \def \showarticletitle #1{#1}   \fi
\ifx \showURL      \undefined \def \showURL       {\relax}        \fi
\providecommand\bibfield[2]{#2}
\providecommand\bibinfo[2]{#2}
\providecommand\natexlab[1]{#1}
\providecommand\showeprint[2][]{arXiv:#2}

\bibitem[\protect\citeauthoryear{Andersen, Borgan, Gill, and Keiding}{Andersen
  et~al\mbox{.}}{1996}]%
        {andersen}
\bibfield{author}{\bibinfo{person}{P.K. Andersen}, \bibinfo{person}{O. Borgan},
  \bibinfo{person}{R. Gill}, {and} \bibinfo{person}{N. Keiding}.}
  \bibinfo{year}{1996}\natexlab{}.
\newblock \bibinfo{booktitle}{\emph{Statistical Models Based on Counting
  Processes}}.
\newblock \bibinfo{publisher}{Springer}.
\newblock


\bibitem[\protect\citeauthoryear{Barrio, Burrage, Leier, and Tian}{Barrio
  et~al\mbox{.}}{2006}]%
        {barrio}
\bibfield{author}{\bibinfo{person}{M. Barrio}, \bibinfo{person}{K. Burrage},
  \bibinfo{person}{A. Leier}, {and} \bibinfo{person}{T. Tian}.}
  \bibinfo{year}{2006}\natexlab{}.
\newblock \showarticletitle{Oscillatory Regulation of hes1: Discrete Stochastic
  Delay Modelling and Simulation}.
\newblock \bibinfo{journal}{\emph{PLoS Computational Biology}}
  \bibinfo{volume}{2}, \bibinfo{number}{9} (\bibinfo{year}{2006}),
  \bibinfo{pages}{1017}.
\newblock


\bibitem[\protect\citeauthoryear{Bouchard-C\^ot\'e, Vollmer, and
  Doucet}{Bouchard-C\^ot\'e et~al\mbox{.}}{2018}]%
        {doucet}
\bibfield{author}{\bibinfo{person}{A. Bouchard-C\^ot\'e},
  \bibinfo{person}{S.~J. Vollmer}, {and} \bibinfo{person}{A. Doucet}.}
  \bibinfo{year}{2018}\natexlab{}.
\newblock \showarticletitle{The Bouncy Particle Sampler: A Non-Reversible
  Rejection Free Markov chain Monte Carlo Method}.
\newblock \bibinfo{journal}{\emph{J. Amer. Statist. Assoc.}}
  \bibinfo{volume}{113} (\bibinfo{year}{2018}), \bibinfo{pages}{855--867}.
\newblock


\bibitem[\protect\citeauthoryear{Br{\'e}maud}{Br{\'e}maud}{1981}]%
        {bremaud1981point}
\bibfield{author}{\bibinfo{person}{P. Br{\'e}maud}.}
  \bibinfo{year}{1981}\natexlab{}.
\newblock \bibinfo{booktitle}{\emph{Point Processes and Queues}}.
\newblock \bibinfo{publisher}{Springer New York}.
\newblock
\showISBNx{9781468494778}
\urldef\tempurl%
\url{https://books.google.fr/books?id=lo-YZwEACAAJ}
\showURL{%
\tempurl}


\bibitem[\protect\citeauthoryear{Bremaud and Massoulie}{Bremaud and
  Massoulie}{1996}]%
        {bremaudStability}
\bibfield{author}{\bibinfo{person}{Pierre Bremaud} {and}
  \bibinfo{person}{Laurent Massoulie}.} \bibinfo{year}{1996}\natexlab{}.
\newblock \showarticletitle{Stability of Nonlinear Hawkes Processes}.
\newblock \bibinfo{journal}{\emph{The Annals of Probability}}
  \bibinfo{volume}{24}, \bibinfo{number}{3} (\bibinfo{year}{1996}),
  \bibinfo{pages}{1563--1588}.
\newblock
\showISSN{00911798}
\urldef\tempurl%
\url{http://www.jstor.org/stable/2244985}
\showURL{%
\tempurl}


\bibitem[\protect\citeauthoryear{Brown, Barbieri, Ventura, Kass, and
  Frank}{Brown et~al\mbox{.}}{2006}]%
        {brown}
\bibfield{author}{\bibinfo{person}{E.N. Brown}, \bibinfo{person}{R. Barbieri},
  \bibinfo{person}{V. Ventura}, \bibinfo{person}{R.E. Kass}, {and}
  \bibinfo{person}{L.M. Frank}.} \bibinfo{year}{2006}\natexlab{}.
\newblock \showarticletitle{The Time-Rescaling Theorem and Its Application to
  Neural Spike Train Data Analysis}.
\newblock \bibinfo{journal}{\emph{Neural Computation}} \bibinfo{volume}{4},
  \bibinfo{number}{2} (\bibinfo{year}{2006}), \bibinfo{pages}{325--346}.
\newblock


\bibitem[\protect\citeauthoryear{Brown, Barbieri, Ventura, Kass, and
  Frank}{Brown et~al\mbox{.}}{2002}]%
        {time-rescale}
\bibfield{author}{\bibinfo{person}{E.~N. Brown}, \bibinfo{person}{R. Barbieri},
  \bibinfo{person}{V. Ventura}, \bibinfo{person}{R.~E. Kass}, {and}
  \bibinfo{person}{L.M. Frank}.} \bibinfo{year}{2002}\natexlab{}.
\newblock \showarticletitle{The Time-Rescaling Theorem and Its Application to
  Neural Spike Train Data Analysis}.
\newblock \bibinfo{journal}{\emph{Neural Computation}} \bibinfo{volume}{14},
  \bibinfo{number}{2} (\bibinfo{year}{2002}), \bibinfo{pages}{325--346}.
\newblock


\bibitem[\protect\citeauthoryear{Cha and Finkelstein}{Cha and
  Finkelstein}{2018}]%
        {cha}
\bibfield{author}{\bibinfo{person}{J.H. Cha} {and} \bibinfo{person}{M.
  Finkelstein}.} \bibinfo{year}{2018}\natexlab{}.
\newblock \bibinfo{booktitle}{\emph{Point Processes for Reliability Analysis}}.
\newblock \bibinfo{publisher}{Springer}.
\newblock


\bibitem[\protect\citeauthoryear{Chevallier, C\'aceres, Doumic, and
  Reynaud-Bouret}{Chevallier et~al\mbox{.}}{2015}]%
        {maria}
\bibfield{author}{\bibinfo{person}{J. Chevallier}, \bibinfo{person}{M.J.
  C\'aceres}, \bibinfo{person}{M. Doumic}, {and} \bibinfo{person}{P.
  Reynaud-Bouret}.} \bibinfo{year}{2015}\natexlab{}.
\newblock \showarticletitle{Microscopic approach of a time elapsed neural
  model}.
\newblock \bibinfo{journal}{\emph{Mathematical Models and Methods in Applied
  Sciences}} \bibinfo{volume}{25}, \bibinfo{number}{14} (\bibinfo{year}{2015}),
  \bibinfo{pages}{2669--2719}.
\newblock


\bibitem[\protect\citeauthoryear{Daley and Vere-Jones}{Daley and
  Vere-Jones}{2003}]%
        {daley}
\bibfield{author}{\bibinfo{person}{Daryl~J Daley} {and} \bibinfo{person}{David
  Vere-Jones}.} \bibinfo{year}{2003}\natexlab{}.
\newblock \bibinfo{title}{An introduction to the theory of point processes.
  Vol. I. Probability and its Applications}.
\newblock
\newblock


\bibitem[\protect\citeauthoryear{Dassios and Zhao}{Dassios and Zhao}{2013}]%
        {dassios}
\bibfield{author}{\bibinfo{person}{A. Dassios} {and} \bibinfo{person}{H.
  Zhao}.} \bibinfo{year}{2013}\natexlab{}.
\newblock \showarticletitle{Exact simulation of Hawkes process with
  exponentially decaying intensity}.
\newblock \bibinfo{journal}{\emph{Electronic Communications in Probability}}
  \bibinfo{volume}{18}, \bibinfo{number}{62} (\bibinfo{year}{2013}).
\newblock


\bibitem[\protect\citeauthoryear{Didelez}{Didelez}{2008}]%
        {didelez}
\bibfield{author}{\bibinfo{person}{V. Didelez}.}
  \bibinfo{year}{2008}\natexlab{}.
\newblock \showarticletitle{Graphical models of markes point processes based on
  local independence}.
\newblock \bibinfo{journal}{\emph{J.R. Statist. Soc. B}} \bibinfo{volume}{70},
  \bibinfo{number}{1} (\bibinfo{year}{2008}), \bibinfo{pages}{245--264}.
\newblock


\bibitem[\protect\citeauthoryear{Doob}{Doob}{1945}]%
        {Doob}
\bibfield{author}{\bibinfo{person}{J.L. Doob}.}
  \bibinfo{year}{1945}\natexlab{}.
\newblock \showarticletitle{Markoff chains – Denumerable case}.
\newblock \bibinfo{journal}{\emph{Trans. Amer. Math. Soc.}}
  \bibinfo{volume}{58}, \bibinfo{number}{3} (\bibinfo{year}{1945}),
  \bibinfo{pages}{455--473}.
\newblock


\bibitem[\protect\citeauthoryear{Gerstner and Kistler}{Gerstner and
  Kistler}{2002}]%
        {gerstner2002spiking}
\bibfield{author}{\bibinfo{person}{W. Gerstner} {and} \bibinfo{person}{W.M.
  Kistler}.} \bibinfo{year}{2002}\natexlab{}.
\newblock \bibinfo{booktitle}{\emph{Spiking neuron models: Single neurons,
  populations, plasticity}}.
\newblock \bibinfo{publisher}{Cambridge university press}.
\newblock


\bibitem[\protect\citeauthoryear{Gillespie}{Gillespie}{1977}]%
        {gillespie}
\bibfield{author}{\bibinfo{person}{D.T. Gillespie}.}
  \bibinfo{year}{1977}\natexlab{}.
\newblock \showarticletitle{Exact Stochastic Simulation of Coupled Chemical
  Reactions}.
\newblock \bibinfo{journal}{\emph{he Journal of Physical Chemistry}}
  \bibinfo{volume}{81}, \bibinfo{number}{25} (\bibinfo{year}{1977}),
  \bibinfo{pages}{2340--2361}.
\newblock


\bibitem[\protect\citeauthoryear{Grazieschi, Leocata, Mascart, Chevallier,
  Delarue, and Tanr\'e}{Grazieschi et~al\mbox{.}}{2019}]%
        {refId0}
\bibfield{author}{\bibinfo{person}{P. Grazieschi}, \bibinfo{person}{M.
  Leocata}, \bibinfo{person}{C. Mascart}, \bibinfo{person}{J. Chevallier},
  \bibinfo{person}{F. Delarue}, {and} \bibinfo{person}{E. Tanr\'e}.}
  \bibinfo{year}{2019}\natexlab{}.
\newblock \showarticletitle{Network of interacting neurons with random synaptic
  weights}.
\newblock \bibinfo{journal}{\emph{ESAIM: ProcS}}  \bibinfo{volume}{65}
  (\bibinfo{year}{2019}), \bibinfo{pages}{445--475}.
\newblock
\urldef\tempurl%
\url{https://doi.org/10.1051/proc/201965445}
\showDOI{\tempurl}


\bibitem[\protect\citeauthoryear{Heger}{Heger}{2004}]%
        {heger}
\bibfield{author}{\bibinfo{person}{D.A. Heger}.}
  \bibinfo{year}{2004}\natexlab{}.
\newblock \showarticletitle{A Disquisition on the Performance Behavior of
  Binary Search Tree Data Structures}.
\newblock  \bibinfo{volume}{5}, \bibinfo{number}{5} (\bibinfo{year}{2004}),
  \bibinfo{pages}{67–75}.
\newblock
\urldef\tempurl%
\url{https://web.archive.org/web/20140327140251/http://www.cepis.org/upgrade/files/full-2004-V.pdf}
\showURL{%
\tempurl}


\bibitem[\protect\citeauthoryear{Herculano-Houzel and Lent}{Herculano-Houzel
  and Lent}{2005}]%
        {Herculano-Houzel2518}
\bibfield{author}{\bibinfo{person}{S. Herculano-Houzel} {and}
  \bibinfo{person}{R. Lent}.} \bibinfo{year}{2005}\natexlab{}.
\newblock \showarticletitle{Isotropic Fractionator: A Simple, Rapid Method for
  the Quantification of Total Cell and Neuron Numbers in the Brain}.
\newblock \bibinfo{journal}{\emph{Journal of Neuroscience}}
  \bibinfo{volume}{25}, \bibinfo{number}{10} (\bibinfo{year}{2005}),
  \bibinfo{pages}{2518--2521}.
\newblock
\showISSN{0270-6474}
\urldef\tempurl%
\url{https://doi.org/10.1523/JNEUROSCI.4526-04.2005}
\showDOI{\tempurl}
\showeprint{https://www.jneurosci.org/content/25/10/2518.full.pdf}


\bibitem[\protect\citeauthoryear{Lewis and Shedler}{Lewis and Shedler}{1978}]%
        {lewis}
\bibfield{author}{\bibinfo{person}{P.A.W. Lewis} {and} \bibinfo{person}{G.S.
  Shedler}.} \bibinfo{year}{1978}\natexlab{}.
\newblock \bibinfo{booktitle}{\emph{Simulation of nonhomogeneous {P}oisson
  processes}}.
\newblock \bibinfo{type}{{T}echnical {R}eport}. \bibinfo{institution}{Naval
  Postgraduate School, Monterey, California}.
\newblock


\bibitem[\protect\citeauthoryear{Mastromatteo, Bacry, and Muzy}{Mastromatteo
  et~al\mbox{.}}{2015}]%
        {mastromatteo}
\bibfield{author}{\bibinfo{person}{Iacopo Mastromatteo},
  \bibinfo{person}{Emmanuel Bacry}, {and} \bibinfo{person}{Jean-Fran\ifmmode
  \mbox{\c{c}}\else~\c{c}\fi{}ois Muzy}.} \bibinfo{year}{2015}\natexlab{}.
\newblock \showarticletitle{Linear processes in high dimensions: Phase space
  and critical properties}.
\newblock \bibinfo{journal}{\emph{Phys. Rev. E}}  \bibinfo{volume}{91}
  (\bibinfo{date}{Apr} \bibinfo{year}{2015}), \bibinfo{pages}{042142}.
\newblock
Issue 4.
\urldef\tempurl%
\url{https://doi.org/10.1103/PhysRevE.91.042142}
\showDOI{\tempurl}


\bibitem[\protect\citeauthoryear{Mesarovic and Takahara}{Mesarovic and
  Takahara}{1975}]%
        {mesarovic}
\bibfield{author}{\bibinfo{person}{M.D. Mesarovic} {and} \bibinfo{person}{Y.
  Takahara}.} \bibinfo{year}{1975}\natexlab{}.
\newblock \bibinfo{booktitle}{\emph{General systems theory: mathematical
  foundations}}. Vol.~\bibinfo{volume}{113}.
\newblock \bibinfo{publisher}{Academic press}.
\newblock


\bibitem[\protect\citeauthoryear{{Muzy}}{{Muzy}}{2019}]%
        {cise-muzy}
\bibfield{author}{\bibinfo{person}{A. {Muzy}}.}
  \bibinfo{year}{2019}\natexlab{}.
\newblock \showarticletitle{Exploiting Activity for the Modeling and Simulation
  of Dynamics and Learning Processes in Hierarchical (Neurocognitive) Systems}.
\newblock \bibinfo{journal}{\emph{Computing in Science Engineering}}
  \bibinfo{volume}{21}, \bibinfo{number}{1} (\bibinfo{date}{Jan}
  \bibinfo{year}{2019}), \bibinfo{pages}{84--93}.
\newblock
\showISSN{1558-366X}
\urldef\tempurl%
\url{https://doi.org/10.1109/MCSE.2018.2889235}
\showDOI{\tempurl}


\bibitem[\protect\citeauthoryear{Muzy and Zeigler}{Muzy and Zeigler}{2014}]%
        {dyndevs}
\bibfield{author}{\bibinfo{person}{Alexandre Muzy} {and}
  \bibinfo{person}{Bernard~P Zeigler}.} \bibinfo{year}{2014}\natexlab{}.
\newblock \showarticletitle{Specification of dynamic structure discrete event
  systems using single point encapsulated control functions}.
\newblock \bibinfo{journal}{\emph{International Journal of Modeling,
  Simulation, and Scientific Computing}} \bibinfo{volume}{5},
  \bibinfo{number}{03} (\bibinfo{year}{2014}), \bibinfo{pages}{1450012}.
\newblock


\bibitem[\protect\citeauthoryear{Muzy, Bacry, Delattre, and M.}{Muzy
  et~al\mbox{.}}{2013}]%
        {Hoff}
\bibfield{author}{\bibinfo{person}{J-F. Muzy}, \bibinfo{person}{E. Bacry},
  \bibinfo{person}{S. Delattre}, {and} \bibinfo{person}{Hoffmann M.}}
  \bibinfo{year}{2013}\natexlab{}.
\newblock \showarticletitle{Modelling microstructure noise with mutually
  exciting point processes}.
\newblock \bibinfo{journal}{\emph{Quantitative Finance}} \bibinfo{volume}{13},
  \bibinfo{number}{1} (\bibinfo{year}{2013}), \bibinfo{pages}{65--77}.
\newblock


\bibitem[\protect\citeauthoryear{Ogata}{Ogata}{1981}]%
        {ogata}
\bibfield{author}{\bibinfo{person}{Y. Ogata}.} \bibinfo{year}{1981}\natexlab{}.
\newblock \showarticletitle{On Lewis' simulation method for point processes}.
\newblock \bibinfo{journal}{\emph{IEEE Transaction on Information Theory}}
  \bibinfo{volume}{27}, \bibinfo{number}{1} (\bibinfo{year}{1981}),
  \bibinfo{pages}{23--31}.
\newblock


\bibitem[\protect\citeauthoryear{Ogata}{Ogata}{1985}]%
        {Ogata_test}
\bibfield{author}{\bibinfo{person}{Y. Ogata}.} \bibinfo{year}{1985}\natexlab{}.
\newblock \showarticletitle{Statistical Models for Earthquake Occurrences and
  Residual Analysis for Point Processes}.
\newblock \bibinfo{journal}{\emph{J. Amer. Statist. Assoc.}}
  \bibinfo{volume}{83}, \bibinfo{number}{401} (\bibinfo{year}{1985}),
  \bibinfo{pages}{9--27}.
\newblock


\bibitem[\protect\citeauthoryear{Pakkenberg, Pelvig, Marner, Bundgaard,
  Gundersen, Nyengaard, and Regeur}{Pakkenberg et~al\mbox{.}}{2003}]%
        {PAKKENBERG200395}
\bibfield{author}{\bibinfo{person}{B. Pakkenberg}, \bibinfo{person}{D. Pelvig},
  \bibinfo{person}{L. Marner}, \bibinfo{person}{M.J. Bundgaard},
  \bibinfo{person}{H.J.G. Gundersen}, \bibinfo{person}{J.R. Nyengaard}, {and}
  \bibinfo{person}{L. Regeur}.} \bibinfo{year}{2003}\natexlab{}.
\newblock \showarticletitle{Aging and the human neocortex}.
\newblock \bibinfo{journal}{\emph{Experimental Gerontology}}
  \bibinfo{volume}{38}, \bibinfo{number}{1} (\bibinfo{year}{2003}),
  \bibinfo{pages}{95 -- 99}.
\newblock
\showISSN{0531-5565}
\urldef\tempurl%
\url{https://doi.org/10.1016/S0531-5565(02)00151-1}
\showDOI{\tempurl}
\newblock
\shownote{Proceedings of the 6th International Symposium on the Neurobiology
  and Neuroendocrinology of Aging.}


\bibitem[\protect\citeauthoryear{Peters and de~With}{Peters and
  de~With}{2012}]%
        {peters}
\bibfield{author}{\bibinfo{person}{E.A.J.F. Peters} {and} \bibinfo{person}{G.
  de With}.} \bibinfo{year}{2012}\natexlab{}.
\newblock \showarticletitle{Rejection-free MonteCarlo sampling for general
  potentials}.
\newblock \bibinfo{journal}{\emph{Physical Review E}} \bibinfo{volume}{85},
  \bibinfo{number}{026703} (\bibinfo{year}{2012}).
\newblock


\bibitem[\protect\citeauthoryear{Reynaud-Bouret, Rivoirard, and
  Tuleau-Malot}{Reynaud-Bouret et~al\mbox{.}}{2013}]%
        {IEEE}
\bibfield{author}{\bibinfo{person}{P. Reynaud-Bouret}, \bibinfo{person}{V.
  Rivoirard}, {and} \bibinfo{person}{C. Tuleau-Malot}.}
  \bibinfo{year}{2013}\natexlab{}.
\newblock \showarticletitle{Inference of functional connectivity in
  Neurosciences via Hawkes processes}. 1st IEEE Global Conference on Signal and
  Information Processing, Austin, Texas.
\newblock


\bibitem[\protect\citeauthoryear{Reynaud-Bouret and Schbath}{Reynaud-Bouret and
  Schbath}{2010}]%
        {RBS}
\bibfield{author}{\bibinfo{person}{P. Reynaud-Bouret} {and} \bibinfo{person}{S.
  Schbath}.} \bibinfo{year}{2010}\natexlab{}.
\newblock \showarticletitle{Adaptive estimation for Hawkes processes;
  application to genome analysis}.
\newblock \bibinfo{journal}{\emph{Annals of Statististics}}
  \bibinfo{volume}{38}, \bibinfo{number}{5} (\bibinfo{year}{2010}),
  \bibinfo{pages}{2781--2822}.
\newblock


\bibitem[\protect\citeauthoryear{Tocher}{Tocher}{1967}]%
        {tocher}
\bibfield{author}{\bibinfo{person}{K.D. Tocher}.}
  \bibinfo{year}{1967}\natexlab{}.
\newblock \bibinfo{booktitle}{\emph{PLUS/GPS III Specification}}.
\newblock \bibinfo{type}{{T}echnical {R}eport}. \bibinfo{institution}{United
  Steel Companies Ltd, Department of Operational Research},
  \bibinfo{address}{Sheffield}.
\newblock


\bibitem[\protect\citeauthoryear{Vere-Jones and Ozaki}{Vere-Jones and
  Ozaki}{1982}]%
        {VeO82}
\bibfield{author}{\bibinfo{person}{D. Vere-Jones} {and} \bibinfo{person}{T.
  Ozaki}.} \bibinfo{year}{1982}\natexlab{}.
\newblock \showarticletitle{Some examples of statistical estimation applied to
  earthquake data.}
\newblock \bibinfo{journal}{\emph{Ann. Inst. Statist. Math.}}
  \bibinfo{volume}{34}, \bibinfo{number}{B} (\bibinfo{year}{1982}),
  \bibinfo{pages}{189--207}.
\newblock


\bibitem[\protect\citeauthoryear{von Bartheld, Bahney, and
  Herculano-Houzel}{von Bartheld et~al\mbox{.}}{2016}]%
        {nbNeurons}
\bibfield{author}{\bibinfo{person}{C.S. von Bartheld}, \bibinfo{person}{J.
  Bahney}, {and} \bibinfo{person}{S. Herculano-Houzel}.}
  \bibinfo{year}{2016}\natexlab{}.
\newblock \showarticletitle{The search for true numbers of neurons and glial
  cells in the human brain: A review of 150 years of cell counting}.
\newblock \bibinfo{journal}{\emph{Journal of Comparative Neurology}}
  \bibinfo{volume}{524}, \bibinfo{number}{18} (\bibinfo{year}{2016}),
  \bibinfo{pages}{3865--3895}.
\newblock


\bibitem[\protect\citeauthoryear{Zeigler}{Zeigler}{1976}]%
        {tms76}
\bibfield{author}{\bibinfo{person}{B.P. Zeigler}.}
  \bibinfo{year}{1976}\natexlab{}.
\newblock \bibinfo{booktitle}{\emph{Theory of Modelling and Simulation}}.
\newblock \bibinfo{publisher}{John Wiley}.
\newblock
\showISBNx{9780471981527}
\showLCCN{75038977}
\urldef\tempurl%
\url{https://books.google.fr/books?id=M-ZQAAAAMAAJ}
\showURL{%
\tempurl}


\bibitem[\protect\citeauthoryear{Zeigler, Muzy, and Kofman}{Zeigler
  et~al\mbox{.}}{2018}]%
        {tms2}
\bibfield{author}{\bibinfo{person}{B.P. Zeigler}, \bibinfo{person}{A. Muzy},
  {and} \bibinfo{person}{E Kofman}.} \bibinfo{year}{2018}\natexlab{}.
\newblock \bibinfo{booktitle}{\emph{Theory of Modeling and Simulation: Discrete
  Event \& Iterative System Computational Foundations}}.
\newblock \bibinfo{publisher}{Academic Press}.
\newblock


\end{thebibliography}
\end{document}